\documentclass[reprint,superscriptaddress,floatfix,longbibliography,tightenlines,nobibnotes,nofootinbib,amsmath,amssymb,aps,pra]{revtex4-1}
\usepackage{graphicx}
\usepackage{color}
\usepackage{bm}
\usepackage{dcolumn}
\usepackage{textcomp}
\usepackage{hyperref}
\hypersetup{colorlinks=true,urlcolor= blue,citecolor=blue,linkcolor= blue,bookmarks=true,bookmarksopen=false}
\usepackage{verbatim}
\usepackage{blindtext}
\usepackage{natbib}
\setcitestyle{numbers,square}
\usepackage{makecell}
\usepackage{amsmath}

\usepackage{amsfonts}
\usepackage{amssymb}
\usepackage{comment}
\usepackage{mathtools}
\usepackage{braket}
\usepackage{soul}
\begin{document}

\title{
Off-resonant light-induced topological phase transition and 
thermoelectric transport in semi-Dirac materials }
\author{Vassilios Vargiamidis}
\email{Vasileios.Vargiamidis@warwick.ac.uk}
\affiliation{School of Engineering, University of Warwick, Coventry, CV4 7AL, United Kingdom}
\author{P. Vasilopoulos }
\email{p.vasilopoulos@concordia.ca}
\affiliation{Department of Physics, Concordia University, 7141 Sherbrooke Ouest, Montreal, Quebec H4B 1R6, Canada}
\author{Neophytos Neophytou}
\email{N.Neophytou@warwick.ac.uk}
\affiliation{School of Engineering, University of Warwick, Coventry, CV4 7AL, United Kingdom}

\begin{abstract}

We show that a semi-Dirac (SD) system with an inversion symmetry breaking mass exhibits a topological phase transition when irradiated with off-resonant light. Using Floquet theory, we derive the band structure, Chern numbers, phase diagram, and we show that as the light intensity is swept at fixed mass, the SD system undergoes normal-Chern-normal insulator transition. Along the phase boundaries we observe single semi-Dirac-cone (SSDC) semimetal states in which one SD cone is gapless and the other gapped. The nontrivial Berry curvature distribution $\Omega(\mathbf{k}) \neq -\Omega ( - \mathbf{k} )$ generates an orbital magnetization $M$ and anomalous Nernst ($\alpha_{xy}$) and thermal Hall ($\kappa_{xy}$) conductivities. We show that $M$ remains constant as the Fermi level  $E_F$ scans the insulating gap, but it changes linearly with it in the Chern insulator (CI) phase, as expected. In the normal insulator phase, we find that $\alpha_{xy}$ exhibits a dip-peak profile which is reversed in the CI phase. We also find that switching the light's circular polarization from left to right 
induces a sign change in $M$, $\alpha_{xy}$, and $\kappa_{xy}$, regardless of the topological phase, thereby allowing us to reverse the direction of flow of the transverse charge and heat currents. Further, we evaluate the components of the charge ($\sigma_{aa}$), thermoelectric ($\alpha_{aa}$), and thermal ($\kappa_{aa}$) conductivity tensors ($a=x, y$) and examine the effect of light on them. With a linear dispersion along the $y$-direction, we find that $\alpha_{yy}$ and $\kappa_{yy}$ are significantly larger than $\alpha_{xx}$ and $\kappa_{xx}$, respectively, due to the much larger squared Dirac velocity $v_y^2$ compared to $v_x^2$.  
	
\end{abstract}

\maketitle
\date{\today}

\section{Introduction}

The realization of topological insulator phases in Dirac materials and of their remarkable properties at their boundaries has been a major challenge \cite{hasan10,qi11}. For instance, the quantum anomalous Hall phase - also called Chern insulator (CI), appears in lattice models with time-reversal symmetry (TRS) breaking Bloch bands carrying a finite Chern number \cite{haldane88} in the absence of a magnetic field. However, it has proved challenging to realize this phase experimentally \cite{chang13}. Another topological phase, the quantum spin Hall insulator (QSHI), originates from spin-orbit coupling and the preservation of TRS \cite{kane05,kane05b,bernevig06}.

Since most materials are intrinsically non-topological, different techniques have been proposed to induce topological phases in them. These include external stimuli, such as pressure \cite{bahramy12}, disorder \cite{groth09,guo10}, impurity doping \cite{xu11}, and heterostructuring combined with electrostatic gating and strain \cite{hasan10}. In the case of the QSHI obtained in materials with a bulk gapped spectrum (e.g. HgTe/CdTe quantum wells), the thickness of the quantum well controls the phase transition \cite{bernevig06b,konig07}. However, most of these techniques require careful band structure engineering and do not exhibit high degree of controllability.

On the other hand, a time-periodic perturbation can be used to manipulate the (Floquet)-Bloch states via the drive frequency, amplitude, and polarization \cite{oka09,kitagawa10,lindner11,platero13,tanaka10,wang13,torres14,torres15,grushin14,narayan16,ebihara16,
tahir15,saha16,fiete18,gumbs22,zubair22,zhou24}, thereby inducing or modifying the topological properties of the bands. In fact, it has been demonstrated that circularly polarized light with suitably chosen parameters opens a gap in semi-metallic systems, allowing them to host gapless edge modes with topological characteristics \cite{kitagawa10,kitagawa11,morell12}.

Some of these exotic effects also appear in another class of two-dimensional (2D) systems, the semi-Dirac (SD) semimetals \cite{pardo09,pick12,huang15} formed with massless and massive dispersions along perpendicular directions. This type of dispersion was found in systems such as the honeycomb lattice in a magnetic field \cite{monta08}, TiO$_2$/V$_2$O$_3$ nanostructures \cite{pardo09b}, photonic crystals \cite{wu14}, surface states of topological insulators \cite{tewari11}, phosphorus-based materials \cite{kim15,prasad16,kim17}, and polariton lattice \cite{real20}, to name a few. SD materials are also promising candidates for valleytronic device applications \cite{ang17}. The dynamic polarization function and plasmons \cite{chackrab16}, transport \cite{nicol19}, and optical \cite{carbot19,peet22} properties in SD materials have also been investigated. It has also been shown that shining light on a semimetallic SD system opens a gap and induces a CI phase with nonzero Chern number \cite{saha16}. However, the topological properties of irradiated SD systems in the presence of a mass term and the consequences for the anomalous thermoelectric transport have not been explored so far.

The purpose of this work is twofold. First, using Floquet theory, we investigate the topological phase transition in a SD system, in the presence of an inversion symmetry breaking mass, irradiated with off-resonant light. 
When  both light and mass are absent, the SD system is semimetallic with two SD nodes, see Fig.~1(a). The presence of the mass term alone opens a trivial gap, see Fig.~1(b). The application of light breaks TRS and opens a topological gap with $\Omega ( - \mathbf{k} ) \neq - \Omega ( \mathbf{k} )$ and finite Chern number. The simultaneous presence of 
mass and light generates different gaps at the two valleys, see Fig.~1(c). In this case we derive the phase diagram and  show that, as the light intensity is swept at fixed mass, the SD system undergoes normal-Chern-normal insulator phase transitions. Along a phase boundary single semi-Dirac-cone (
SSDC) states appear, in which one SD cone is gapless and the other  gapped, cf. Sec.~III. We also investigate the fate of the CI phase at finite temperatures, when the system is in a quantum mixed state described by a density matrix. We find that the CI phase survives for temperatures $\lesssim 30$ K for experimentally realizable parameters.
\begin{figure*}[t]
\vspace*{0.2cm}
\begin{center}
\includegraphics[height=9cm, width=16cm ]{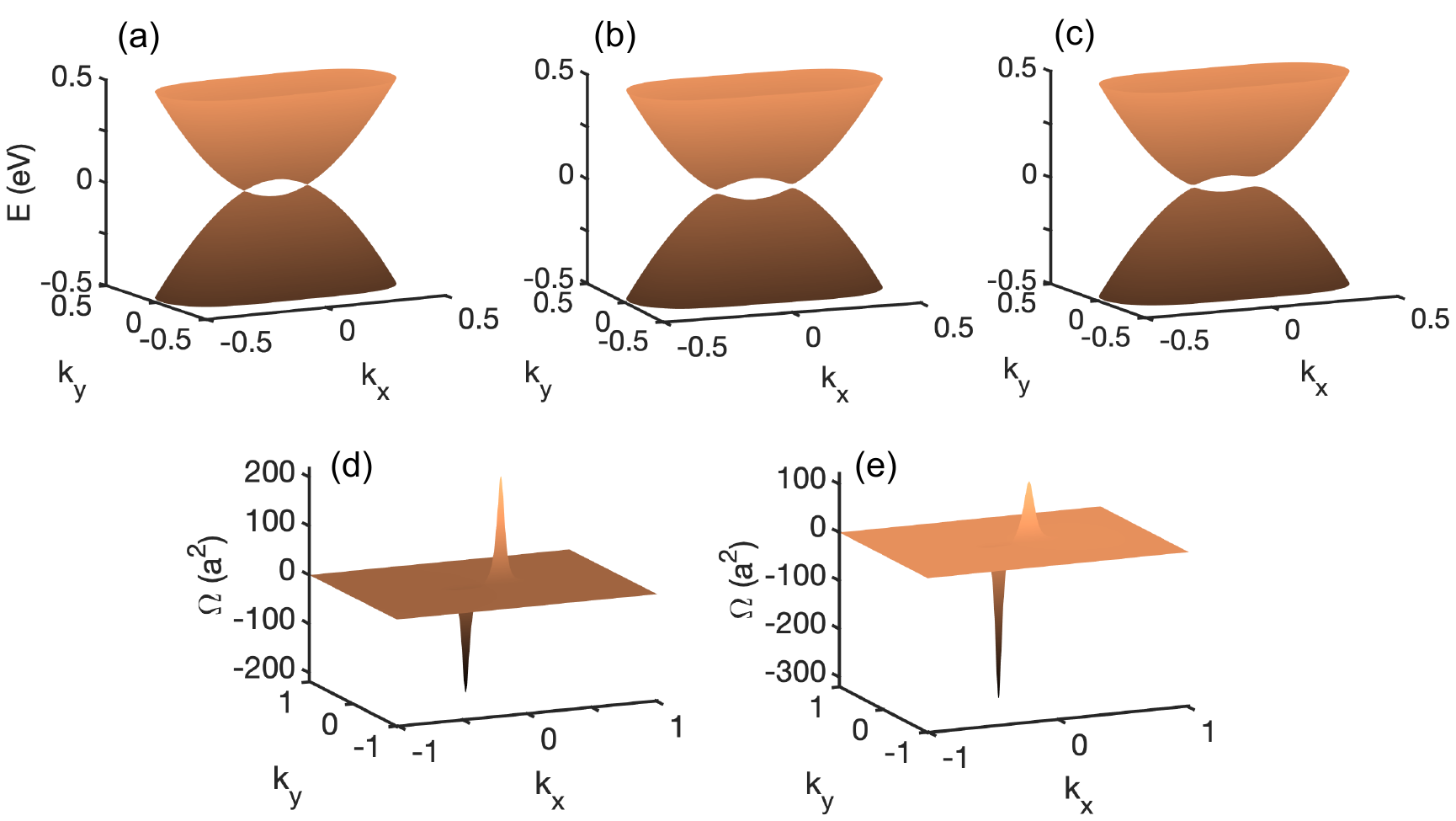}
\end{center}
\vspace*{-0.3cm} \caption{(Colour online) (a)-(c) The energy spectrum of a SD material with a gap parameter $\Delta = - 40$ meV. In (a) we plot the spectrum for $\tilde{A} = m = 0$; the system is semi-metallic with two gapless SD nodes. In (b) $\tilde{A} = 0$ and $m = 10$ meV; the mass term gaps the SD nodes. The corresponding Berry curvature is shown in Fig.~1(d). It satisfies $\Omega ( - \mathbf{k} ) = - \Omega( \mathbf{k} )$ due to the preservation of TRS. In (c) $\tilde{A} = 0.09$ \AA$^{-1}$, $\omega = 3 \times 10^{15}$ Hz, and $m = 10$ meV; the simultaneous presence of light and mass generates different gaps at the two valleys. In this case, $\Omega ( - \mathbf{k} ) \neq - \Omega( \mathbf{k} )$ as a consequence of the broken TRS, see Fig.~1(e).   }
\label{fig:fig1}%
\end{figure*} 

The nonzero local Berry curvature in irradiated SD materials induces anomalous transport responses, such as the anomalous Hall effect (AHE), the anomalous Nernst effect (ANE), and the anomalous thermal Hall effect (ATHE). It also generates an orbital magnetization (OM). Our second objective is to investigate these responses in the (normal) band insulator (BI) and CI phases. The OM, a quantity closely related to the ANE and AHE, remains constant as 
$E_F$ scans the insulating gap, but it changes linearly with it in the CI phase. 
Importantly, we find that switching the light's circular polarization,  from left to right, induces a sign change in $M$ and the transverse conductivities $\sigma_{xy}$, $\alpha_{xy}$ and $\kappa_{xy}$. 

Further, we evaluate $\sigma_{xx}$ and $\sigma_{yy}$ and find analytical expressions for zero temperature when the spectrum exhibits a gapless SD point. We also find that the effect of light on $\alpha_{xx}$ is opposite to that on $\alpha_{yy}$; $\alpha_{xx}$ decreases with the light intensity whereas $\alpha_{yy}$ decreases. The thermal conductivities $\kappa_{xx}$ and $\kappa_{yy}$ are also evaluated; they exhibit a behaviour similar to those of $\sigma_{xx}$ and $\sigma_{yy}$.

In Sec.~II we derive an effective static (Floquet) Hamiltonian, the energy spectrum, and Berry curvature of an irradiated SD material with a mass term. In Sec.~III we derive the Chern numbers, phase diagram, and discuss the phase transitions. The OM is evaluated in Sec.~IV and the anomalous thermoelectric transport in Sec.~V. We summarize and conclude in Sec.~VI.

\section{Floquet Theory and Berry curvature}

We consider a SD material under the application of a time-dependent radiation generated by an off-resonant polarized light field. The driven system is described by a time-periodic Hamiltonian $H(\mathbf{k}, t) = H(\mathbf{k}, t+T)$, where $T = 2 \pi / \omega$ is the driving period and $\omega$ the driving frequency. The time dependence of the electric field is induced via the electromagnetic vector potential, $\mathbf{A} ( t ) = ( A_x (t), A_y (t) )$ which is taken to be
\begin{equation}
\mathbf{A}(t) = A_0 [ \cos ( \omega t ) \hat{x} + \cos ( \omega t - \phi ) \hat{y} ]  ,
\label{eq1}%
\end{equation}
where $\phi$ allows tuning the light polarization and $A_0 = E_0 / \omega$ is the driving amplitude with $E_0$ the amplitude of the electric field $\mathbf{E} = - \partial \mathbf{A} / \partial t$. For $\phi = \pi / 2$ ($-\pi / 2$) the light is left (right) circularly polarized; for $\phi = 0$ and $\pi$ it is linearly polarized. 

The coupling to the electromagnetic field is introduced 
by replacing the momentum $\mathbf{p}$ with the covariant momentum $\mathbf{p} - e \mathbf{A}(t)$, where $e<0$ is the electron charge. We assume that the sample is smaller than the wavelength of light so there is no spatial dependence of the driving field. Then the irradiated SD material is described by the time-dependent Bloch Hamiltonian
\begin{equation}
H ( \mathbf{k}, t ) = \left[ \Delta + \alpha \pi_x^2(t) \right] \sigma_x + \gamma \pi_y(t) \sigma_y + m \sigma_z  ,
\label{eq2}%
\end{equation}
where $\pi_j(t) = k_j - ( e/\hbar ) A_j(t)$ with $j=x,y$; $\Delta$ is the gap parameter \cite{goerbig09,fuch09}, $m$ is the mass term, $\gamma = \hbar v_y$ with $v_y$ the Dirac velocity along the $y$ direction, $\alpha = \hbar^2 / 2 m^\star$ with $m^\star$ the effective mass, associated with the $x$ direction, and $\boldsymbol{\sigma} = ( \sigma_x, \sigma_y, \sigma_z )$ the Pauli matrices in the sublattice pseudospin space. We use $m^\star \simeq 1.49 m_e$ where $m_e$ is the free electron mass and $v_y \simeq 3 \times 10^5$ m/s, which are parameters relevant to monolayer black phosphorus \cite{kim15}.

In the irradiated SD material, a quantum state evolves as $\vert \psi_{\mathbf{k}} \rangle = U_{\mathbf{k}}(t, t_0) \vert \psi_{\mathbf{k}} (t, t_0) \rangle$ where $U_{\mathbf{k}} (t, t_0) = \mathcal{T} \exp \left( - (i / \hbar ) \int_{t_{0}}^{t} H(\mathbf{k}, t^{\prime}) d t^{\prime} \right)$ is the time-evolution operator with $\mathcal{T}$ the time-ordering operator and $t_0$ is the initial time of the perturbation. For photon energies much larger than that of the electrons, $H ( \mathbf{k}, t )$ can be reduced to an effective time-independent Hamiltonian using Floquet theory \cite{kitagawa11}. We define 
this Hamiltonian through the time-evolution operator over one period $U_{\mathbf{k}}(T, 0)$ as
\begin{equation}
\mathcal{T} \exp \bigg( - \frac{i}{\hbar} \int_{0}^{T} H(\mathbf{k}, t^{\prime}) d t^{\prime} \bigg) = \exp \bigg( - \frac{i}{\hbar} H_F ( \mathbf{k} ) T \bigg)   ,
\label{eq3}%
\end{equation}
where we have assumed $t_0 = 0$. A high-frequency expansion of $U_{\mathbf{k}}(T, 0)$, up to first order in inverse frequency, yields the Floquet Hamiltonian
\begin{equation}
H_F ( \mathbf{k} ) \simeq H_0 + \frac{[ H_{-1}, H_{+1} ]}{\hbar \omega}  ,
\label{eq4}%
\end{equation}  
where $H_n = ( 1 / T) \int_0^T e^{- i n \omega t} H ( \mathbf{k}, t ) d t$ is the discrete Fourier component of the time-periodic Hamiltonian $H( \mathbf{k}, t )$. This approach is valid for high frequencies and low intensities, such that $e A_0 v_y \ll \hbar \omega$. The off-resonant light does not excite electrons and there is no optical absorption. Instead, it modifies the electron band structure through second-order virtual photon absorption processes that
lead to the creation or modification of the band gap \cite{kitagawa11,morell12}.

Using Eqs.~(\ref{eq2})-(\ref{eq4}) we obtain 
\begin{equation}
H_0 = \left( \Delta + \alpha k_x^2 + \frac{\alpha e^2 A_0^2}{2 \hbar^2} \right) \sigma_x + \gamma k_y \sigma_y + m \sigma_z   ,
\label{eq5}%
\end{equation}
and
\begin{equation}
H_{\pm 1} = - \frac{1}{\hbar} \alpha e A_0 k_x \sigma_x - \frac{1}{2} e A_0 v_y e^{\mp i \phi} \sigma_y  .
\label{eq6}%
\end{equation}    
Evaluating the commutator in Eq.~(\ref{eq4}) gives $[ H_{-1}, H_{+1} ] = \hbar \omega \eta k_x \sigma_z$ where $\eta = 2 \alpha v_y \tilde{A}^2 \sin \phi / \omega$ with $\tilde{A} = e A_0 / \hbar$. Then the Floquet Hamiltonian reads
\begin{equation}
H_F ( \mathbf{k} ) = \mathbf{d} ( \mathbf{k} ) \cdot \boldsymbol{\sigma}  ,
\label{eq7}%
\end{equation}  
where
\begin{equation}
\mathbf{d} ( \mathbf{k} ) = ( \Delta_1 + \alpha k_x^2, \gamma k_y, m_z )  ,
\label{eq8}%
\end{equation} 
with $\Delta_1 = \Delta + \alpha \tilde{A}^2 / 2$ and $m_z = m+\eta k_x$. The energy dispersion, obtained by squaring the Hamiltonian, is $E_{\lambda \mathbf{k}} = \sqrt{\mathbf{d}( \mathbf{k} ) \cdot \mathbf{d} ( \mathbf{k} ) } = \lambda \varepsilon_{\mathbf{k}}$ with $\lambda = +1 (-1)$ for the conduction (valence) band and
\begin{equation}
\varepsilon_{\mathbf{k}} = \big[ \left( \Delta_1 + \alpha k_x^2 \right)^2 + \gamma^2 k_y^2 + ( m + \eta k_x )^2 \big]^{1/2}   .
\label{eq9}%
\end{equation}  
In the above model, the light-induced term $h( \mathbf{k} ) = \eta k_x \sigma_z$ breaks TRS; explicitly, $\Theta h( \mathbf{k} ) \Theta^{-1} \neq h( - \mathbf{k} )$, with $\Theta = \mathcal{K}$ the complex conjugation operator. For linearly polarized light ($\phi = 0, \pi$), for which $\eta = 0$, TRS is preserved.

In Figs.~1(a)-1(c) we show the energy dispersion $E_{\lambda \mathbf{k}}$ as a function of $\mathbf{k}$ for $\Delta = - 40$ meV. In Fig.~1(a), $\tilde{A} = m =0$, the system is in the semi-metallic state with two gapless SD nodes located at $\mathbf{k}_D^{\pm} = \left( \pm \sqrt{- \Delta / \alpha}, 0 \right)$ \cite{remark}. In Fig.~1(b), $\tilde{A} =0$ and $m = 10$ meV, TRS is preserved and the band gap is trivial. The Berry curvature evaluated in Eq.~(\ref{eq11}) is shown in Fig.~1(d). The spectrum under the application of light and in the presence of mass is shown in Fig.~1(c) and the Berry curvature in Fig.~1(e) for $\tilde{A} = 0.09$ \AA$^{-1}$, $\omega = 3 \times 10^{15}$ Hz, and $m = 10$ meV. In this case $\Delta_1 < 0$, the light gaps the SD nodes and shifts their positions at $\mathbf{k}_D^{\pm} = \left( k_D^{\pm}, 0 \right)$ with $k_D^{\pm} = \pm \sqrt{ - \Delta_1  / \alpha}$ \cite{dirac}. As will be shown below, when $\Delta_1 < 0$ the system is a CI for $| m | < | \eta k_D |$ and a BI for $| m | > | \eta k_D |$. We remark that for $\Delta_1 = 0$, i.e., for $\Delta = - \alpha \tilde{A}^2 / 2$, it is evident from Eq.~(\ref{eq9}) that the spectrum exhibits a gapless (gapped) SD point for $m = 0$ ($m \neq 0$). For $\Delta_1 > 0$ the system is a BI regardless the value of $m$, as will be shown later. 

\textit{Berry curvature}. For the above two-band model, the Berry curvature can be expressed as \cite{hasan10}
\begin{equation}
\Omega ( \mathbf{k} ) = \frac{1}{2 | \mathbf{d} |^3} \mathbf{d} \cdot \left( \partial_{k_{x}} \mathbf{d} \times \partial_{k_{y}} \mathbf{d} \right)   .
\label{eq10}%
\end{equation}  
Using $\mathbf{d} ( \mathbf{k} )$ from Eq.~(\ref{eq8}) we get
\begin{equation}
\Omega ( \mathbf{k} ) = \frac{\gamma \eta ( \alpha k_x^2 - \Delta_1 ) + 2 \alpha \gamma m k_x}{2 \varepsilon_{\mathbf{k}}^3}   .
\label{eq11}%
\end{equation}
In the absence of light, $\eta = 0$, Eq.~(\ref{eq11}) satisfies $\Omega ( - \mathbf{k} ) = - \Omega ( \mathbf{k} )$. Under the application of light, $\eta \neq 0$, it satisfies $\Omega ( - \mathbf{k} ) \neq - \Omega ( \mathbf{k} )$.    
\begin{figure*}[t]
\vspace*{0.2cm}
\begin{center}
\includegraphics[height=10cm, width=17cm ]{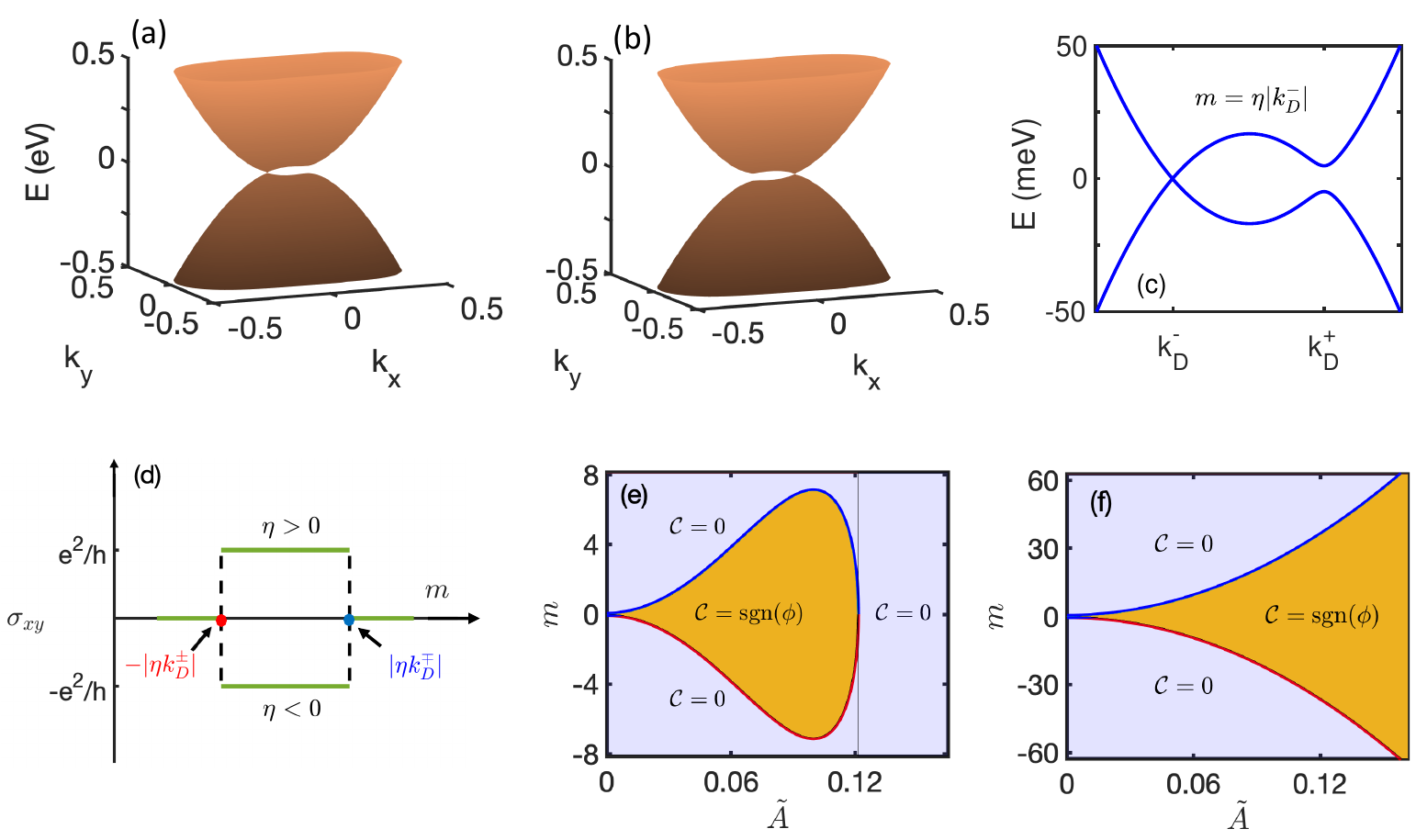}
\end{center}
\vspace*{-0.3cm} \caption{(Colour online) (a)-(b) The band structure of a SD material in the SSDC semimetal states at the phase transition points (a) $m = \eta k_D$ and (b) $m = - \eta k_D$. The parameter values are $\Delta = - 20$ meV, $\tilde{A} = 0.09$ \AA$^{-1}$, $\omega = 3 \times 10^{15}$ Hz, and $\phi = \pi / 2$. In (a) the gap is open at $k_D^{+}$ but closed at $k_D^{-}$ with linear dispersion along $k_y$. (c) The band structure of (a) for $k_y = 0$ showing the SSDC state. (d) $\sigma_{xy}$ versus $m$ for both left- and right-circularly polarized light. (e) $(m, \tilde{A})$ phase diagram. The thin vertical line at $\tilde{A}_{th} \simeq 0.12$ \AA$^{-1}$ corresponds to $\Delta_1 = 0$. (f) $(m, \tilde{A})$ phase diagram for an isotropic model with linear bands.   }
\label{fig:fig1}%
\end{figure*} 

\section{Chern numbers and topological transition}

We characterize the topology of the irradiated SD system by explicitly calculating the Chern numbers. We employ the method of the Brouwer degree \cite{simon12} in order to evaluate the Chern number as
\begin{equation}
\mathcal{C} = \frac{1}{2} \sum_{\mathbf{k} \in \mathbf{D}_i } \text{sgn}(J_z)_i \text{sgn}(d_z)_i  ,
\label{eq12}%
\end{equation}  
where $J_z =  \partial_{k_{x}} d_x \partial_{k_{y}} d_y - \partial_{k_{x}} d_y \partial_{k_{y}} d_x$ is the Jacobian and $\text{sgn}(x)$ is the sign function. In this approach, instead of integrating the Berry curvature in the Brillouin zone, we need to evaluate a discrete sum over the momenta $\mathbf{k}$ in the set of Dirac points $\mathbf{D}_i$. The Jacobian is found to be $J_z = 2 \alpha \gamma k_x$; at the Dirac points $\text{sgn}\left[ J_z ( \mathbf{k}_D^{\pm} ) \right] = \pm 1$, which also gives their chirality. The sign of the mass term at the Dirac points is expressed as $\text{sgn}\left[m_z ( \mathbf{k}_D^{\pm} )\right] = \text{sgn}( m \pm \eta k_D)$ with $k_D = | k_D^{+} | = | k_D^{-} |$. Thus the Chern number reads
\begin{equation}
\mathcal{C} = \frac{1}{2} \left[ \text{sgn} ( m + | \eta | k_D ) - \text{sgn} ( m - | \eta | k_D ) \right]   .
\label{eq13}%
\end{equation}  
Assuming the polarization of light is $0 < \phi < \pi$ ($\eta > 0$), we find $\mathcal{C} = 0$ for $m > \eta k_D$; the mass term has the same sign at the Dirac points. As $m$ derceases (or as the light intensity increases), the Dirac node at $k_D^{-}$ goes through a gap closing and opening transition at $m = \eta | k_D^{-} |$; the gap at the second Dirac node stays open, see Fig.~2(a). For $- \eta k_D < m < \eta k_D$ we find $\mathcal{C} = 1$; the SD system is in the CI phase. As $m$ further decreases, the second Dirac node undergoes a gap closing and opening transition at $m = - \eta k_D^{+}$; the first Dirac node remains gapped, see Fig.~2(b). For $m < - \eta k_D$, the system is a BI with $\mathcal{C} = 0$. When the polarization is $- \pi < \phi < 0$ ($\eta < 0$), the order of the gap closings is reversed. In this case we find $\mathcal{C} = - 1$ for $- | \eta | k_D < m < | \eta | k_D$, and $\mathcal{C} = 0$ otherwise. These results can be expressed as 
\begin{equation}
  \mathcal{C}=\left\{
  \begin{array}{@{}ll@{}}
    \text{sgn} ( \phi ), &  | m | < | \eta | k_D \\
    0, & | m | > | \eta | k_D
  \end{array} \right. 
\label{eq14}%
\end{equation} 
The SSDC states appear at the phase transition points, $| m | = | \eta | k_D$, where only one SD cone is gapless and the other one is gapped. In Fig.~2(c) the SSDC state is shown for $m = \eta | k_D^{-} |$; it was obtained from the band structure of Fig.~2(a) by plotting $E_{\lambda \mathbf{k}}$ versus $k_x$ for $k_y = 0$. These remarkable states can only be realized when light and mass are simultaneously present. They were found previously in photoirradiated silicene \cite{ezawa13} and recently in jacutingaite \cite{vargiam22} without the application of light but with broken inversion and TRS.

In Fig.~2(d) we show the Hall conductivity $\sigma_{xy}$ as a function of the mass $m$ for $\eta > 0$ and $\eta < 0$. For $| m | < | \eta | k_D$, $\sigma_{xy} = ( e^2 / h ) \text{sgn}(\eta)$; $\sigma_{xy} = 0$ otherwise. At the phase transition points, $\sigma_{xy}$ jumps from $0$ to $\pm e^2 / h$.

In Fig.~2(e) we show the $(m, \tilde{A})$ phase diagram. The blue (red) line corresponds to the gap closing and opening at $| k_D^{-} |$ ($k_D^{+}$) for $\eta > 0$; for $\eta < 0$ they are swapped. These lines correspond to the SSDC states. We notice that the SD system undergoes normal-Chern-normal insulator phase transition as the light intensity increases from zero at fixed value of the mass. As $\tilde{A}$ approaches the threshold value $\tilde{A}_{th} \simeq 0.12$ \AA$^{-1}$ for which $\Delta_1 \rightarrow 0$ (see thin vertical line in Fig.~2(e)), the SSDC state gradually shifts and becomes a SD node at $\mathbf{k}_D = 0$. For $\tilde{A} > \tilde{A}_{th}$ so that $\Delta_1 > 0$, the SD node is gapped; in this case the system is a BI with $\mathcal{C} = 0$ \cite{dirac2}. These transitions also occur when varying $m$ at fixed value of $\tilde{A}$. We emphasize here that for $m = 0$, there are no SSDC states and the gaps are the same at the two valleys; in this case $\mathcal{C} = \text{sgn}( \phi )$ for $\Delta_1 < 0$ and $\mathcal{C} = 0$ for $\Delta_1 > 0$ \cite{saha16}.

\textit{Comparison with isotropic Dirac systems.} It is instructive to make a comparison with light-induced topological transitions in isotropic Dirac systems possessing linear bands. A prototype Dirac system is the honeycomb lattice (for example monolayer graphene), whose low-energy Hamiltonian is given by
\begin{equation}
H_g ( \mathbf{k} ) = \hbar v_F ( \tau_z \sigma_x k_x + \sigma_y k_y ) + m \sigma_z ,
\label{eq150}%
\end{equation}
where $\tau_z = \pm 1$ is the valley pseudospin for the $K$ and $K^{\prime} = - K$ valleys. We have included an inversion symmetry breaking mass $m$ originating from a site energy difference $2 m$ between sublattices. When irradiated with off-resonant light, we can use the vector potential in Eq.~(\ref{eq1}) and the Floquet approach to obtain an effective time-independent Hamiltonian
\begin{equation}
H_F^g ( \mathbf{k} ) = \mathbf{d}_g ( \mathbf{k} ) \cdot \boldsymbol{\sigma} ,
\label{eq151}%
\end{equation}
where
\begin{equation}
\mathbf{d}_g ( \mathbf{k} ) = ( \hbar v_F \tau_z k_x, \hbar v_F k_y, m_g ) ,
\label{eq152}%
\end{equation}
with $m_g = m + \eta^{\prime} \tau_z$ and $\eta^{\prime} = \hbar v_F^2 \tilde{A}^2 \sin \phi / \omega$. The energy dispersion is given by $E_{\lambda \mathbf{k}} = \lambda \varepsilon_{\mathbf{k}}^{\prime}$ with $\varepsilon_{\mathbf{k}}^{\prime} = ( \hbar^2 v_F^2 k^2 + m_g^2 )^{1/2}$. The light-induced term $h_{\tau_{z}} ( \mathbf{k} ) = \eta^{\prime} \tau_z \sigma_z$ is of the Haldane type and breaks TRS; explicitly, $\Theta h_{\tau_{z}} ( \mathbf{k} ) \Theta^{-1} \neq h_{\tau_{z}} ( - \mathbf{k} )$, with $\Theta = \tau_x \mathcal{K}$. For $\eta^{\prime} > 0$, this term is negative at $K^{\prime}$ and positive at $K$; the edge states will connect one Dirac cone with another by crossing the gap \cite{graphene}. The light-induced term in SD systems, $h ( \mathbf{k} ) = \eta k_x \sigma_z$, is negative (positive) at $k_x = - k_D ( + k_D )$ and the behaviour of the edge states will be similar. 

In order to evaluate the Chern number we need the sign of the Jacobian at the Dirac points $\mathbf{k}_D^{\pm} = \pm \mathbf{K}$. We find $J_z = \hbar^2 v_F^2 \tau_z$ and at the Dirac points $\text{sgn} [ J_z ( \mathbf{k}_D^{\pm} ] = \pm 1$. The mass sign at the Dirac points is $\text{sgn} [ m_g ( \mathbf{k}_D^{\pm} ) ] = \text{sgn} ( m \pm \eta^{\prime})$. Thus the Chern number reads
\begin{equation}
\mathcal{C} = \frac{1}{2} \left[ \text{sgn} ( m + |\eta^{\prime}| ) - \text{sgn} ( m - |\eta^{\prime}| ) \right]  ,
\label{eq154}%
\end{equation}
As $m$ varies we obtain gap closing and reopening transitions at $m = \pm | \eta^{\prime} |$ similarly to SD systems; the mass term vanishes at one Dirac point, $m + \eta^{\prime} \tau_z = 0$. For $m = \eta^{\prime}$ the Dirac cone at $K^{\prime}$ is gapless and at $K$ is gapped. For $m= - \eta^{\prime}$ they are swapped.

However, there is an important difference between the Chern numbers in Eqs.~(\ref{eq13}) and (\ref{eq154}). An isotropic Dirac system remains gapped with $\mathcal{C} = \text{sgn} ( \phi )$ regardless of the light intensity $\tilde{A}$ as long as $| m | < | \eta^{\prime}|$. This is shown in the $( m, \tilde{A} )$ phase diagram in Fig.~2(f). The blue (red) line corresponds to the gap closing at the $K^{\prime}$ ($K$) valley for $\eta^{\prime} > 0$; for $\eta^{\prime} < 0$ they are swapped. The Dirac system undergoes normal to Chern insulator phase transition as $\tilde{A}$ increases from zero; one can traverse the phase boundary between the regions with $\mathcal{C} = 0$ and $\mathcal{C} = \text{sgn} ( \phi )$ only once. This is in contrast to a SD system which undergoes normal-Chern-normal phase transition as $\tilde{A}$ increases from zero. As $\tilde{A}$ reaches $\tilde{A}_{th}$ ($\Delta_1 \rightarrow 0$) the topological gap closes and reopens with $\mathcal{C} = 0$; one can traverse the phase boundary twice, see Fig.~2(e). This is due to the presence of $k_x$ in the mass term $m_z$ in Eq.~(\ref{eq8}) which becomes $k_D$ in the Chern number in Eq.~(\ref{eq13}). 

\textit{Anomalous Hall conductivity.} Using the Berry curvature from Eq.~(\ref{eq11}) we evaluate the anomalous Hall conductivity as
\begin{equation}
\sigma_{xy} = \frac{e^2}{h} \int \frac{d^2 k}{( 2 \pi )^2} \left( f_{- \mathbf{k}} - f_{+ \mathbf{k}} \right) \Omega ( \mathbf{k} )  ,
\label{eq15}%
\end{equation}
where $f_{\pm \mathbf{k}} = \{ 1 + \exp \left[ \beta ( \pm \varepsilon_{\mathbf{k}} - E_F ) \right] \}^{-1}$ is the Fermi distribution function of the conduction ($+$) and valence ($-$) bands and $\beta = 1 / k_B T$ with $k_B$ the Boltzmann constant and $T$ the temperature. In Fig.~3 we show numerical results of $\sigma_{xy}$ versus $E_F$ for left-circularly polarized light ($\phi = \pi / 2$) and $T = 0.1$ K. We have used $\Delta = - 20$ meV, $\omega = 3 \times 10^{15}$ Hz, and $m = 2.5$ meV. The values of $\tilde{A}$ are such that $\Delta_1 < 0$ in both Fig.~3(a) and 3(b) \cite{comment}. In Fig.~3(a) we observe that $\sigma_{xy}$ is finite and quantized for $E_F$ in the gap; it is in the CI phase. As the light intensity decreases the energy gap gradually shrinks and brings the system to a SSDC state where the gap closes at $| m | = | \eta | k_D$. It then reopens for $| m | > | \eta | k_D$ and further decrease of the light intensity $\tilde{A}$ causes the AHC to vanish inside the gap, as shown in Fig.~3(b). Thus the interplay of light and mass (see Ref.~\cite{mass} and Appendix A) offers the possibility to change the topological phase. We also notice that, away from the gap, the Hall conductivity is small. This is due to the small value of $\eta$ for weak light intensity $\tilde{A}$.

In Fig.~3(c) we show $\sigma_{xy}$ versus $E_F$ in the CI phase for left- and right-circularly polarized light, $\phi = \pi / 2$ and $ - \pi / 2$, respectively. We notice that $\sigma_{xy}$ is reversed when switching the polarization of light indicating that the direction of the edge state is reversed; this is also evident from Eq.~(\ref{eq14}).
\begin{figure}[t]
\vspace*{0.0cm}
\begin{center}
\hspace{-0.2cm}\includegraphics[height=7.8cm, width=8.6cm ]{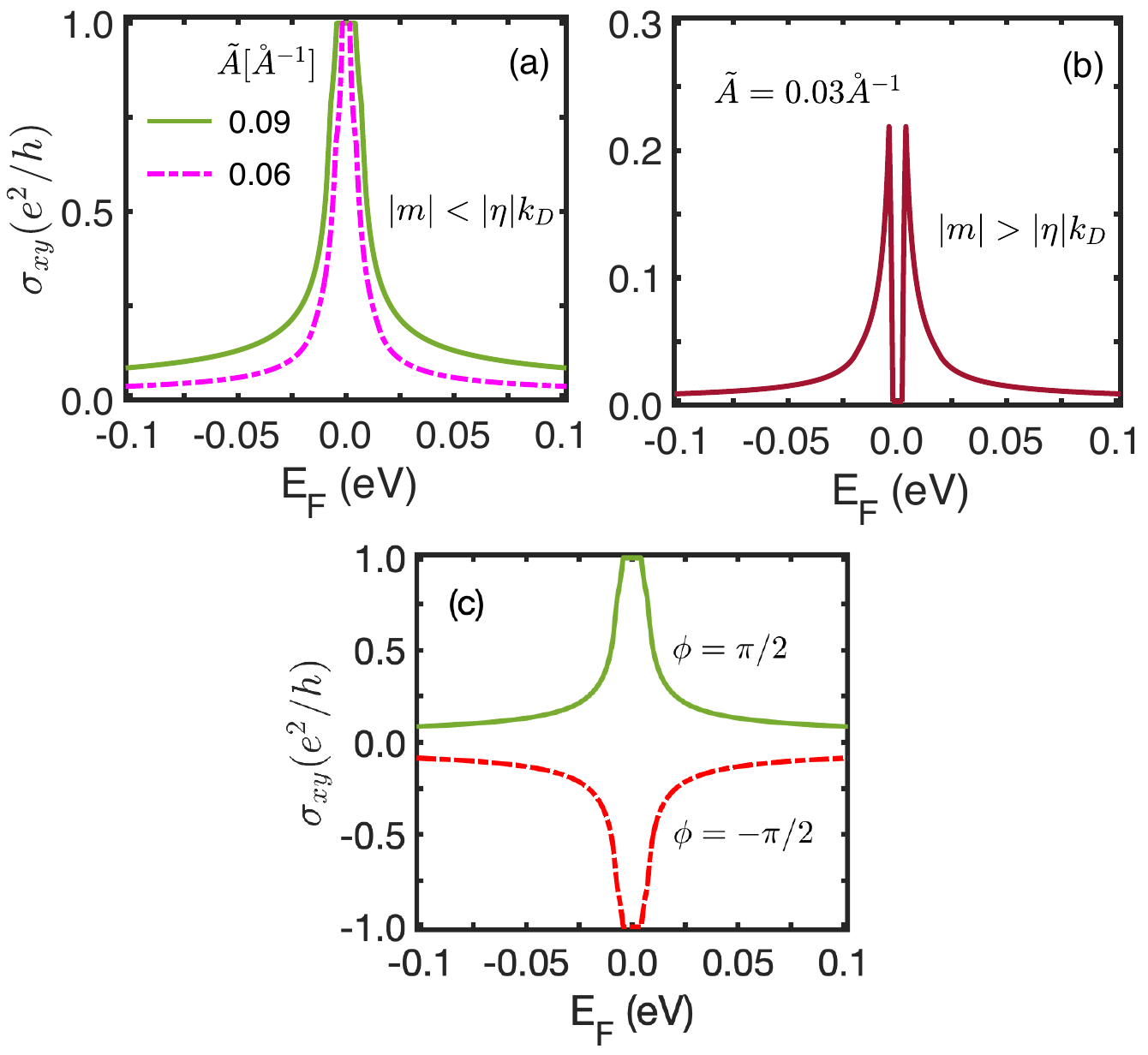}
\end{center}
\vspace*{-0.4cm} \caption{(Colour online) (a)-(b) Anomalous Hall conductivity $\sigma_{xy}$ as a function of $E_F$ for $\Delta_1 < 0$. In (a) the SD system is a CI with $\sigma_{xy} = e^2 / h$ for $E_F$ in the gap and in (b) it is a BI with $\sigma_{xy} = 0$; the topological phase transition occurs at $m = \
| \eta | k_D$. (c) $\sigma_{xy}$ versus $E_F$ for left- and right-circularly polarized light in the CI phase.   }
\label{fig:fig1}%
\end{figure}

\textit{Temperature effects on the CI phase.} Chern invariants for systems with broken TRS lead to quantized Hall conductivities at temperatures well below the energy gap. However, at finite temperatures, quantization is lost due to thermal averaging or dephasing to mixed states \cite{rivas13,arovas14,rivas14,vander16,guo18}. This issue has lead to the question of a proper definition of a topological invariant for a mixed rather than a pure state. The extension of the Chern invariant based on the density matrix for mixed states is rather nontrivial and no universally agreed definition exists. Here, we employ the mean Uhlmann curvature (MUC) \cite{car18} to evaluate the thermal Uhlmann number.

In thermal equilibrium, the state of the system is described by the density operator $\rho = e^{- \beta H_F} / \mathcal{Z}$, where $\mathcal{Z} = \text{tr} ( e^{- \beta H_F} )$ is the partition function, with $H_F$ given in Eq.~(\ref{eq7}); it can be readily evaluated to give $\mathcal{Z} = 2 \cosh ( \beta | \mathbf{d} | )$. For a two-band model, the MUC, which is a mixed state generalization of the Berry curvature, is given as \cite{guo18,car18}
\begin{equation}
\mathcal{U} ( \mathbf{k}, T ) = \frac{\tanh^3 ( \beta | \mathbf{d} | )}{2 | \mathbf{d} |^3} \mathbf{d} \cdot ( \partial_{k{x}} \mathbf{d} \times \partial_{k{y}} \mathbf{d} )  .
\label{eq1600}%
\end{equation}
In the limit $T \rightarrow 0$, $\mathcal{U} ( \mathbf{k}, T )$ reduces to the Berry curvature in Eq.~(\ref{eq10}). For $T \rightarrow \infty$ it vanishes; the valence and conduction bands are equally populated in this limit, but they contribute with opposite curvatures. In a continuum model Hamiltonian, the Uhlmann number is calculated by integrating the MUC in the entire momentum space:
\begin{equation}
n_{U} = \frac{1}{2 \pi} \int \mathcal{U} ( \mathbf{k}, T ) d^2 k  .
\label{eq1601}%
\end{equation}

In Fig.~4(a) we show $n_U$ as a function of temperature $T$ for $E_F$ placed in the middle of the gap. The values of the light intensity and of the mass are such that the system is in the CI phase for $T = 0$; it is inside the topological region in Fig.~2(e). Notice that for $T = 0$, $n_U$ correctly reproduces the Chern number result. The system remains in the CI phase, $n_U = 1$, even at nonzero temperature, provided that it is below a certain critical $T_c$ associated with the band gap. For the chosen parameters, $n_U$ starts losing its quantized value at $T_c \simeq 30$ K for which $k_B T \simeq 2.6$ meV. This is comparable to the band gap ($\simeq 5$ meV). It is reasonable to expect that, for larger topological gap, the CI phase remains stable ($n_U = 1$) for larger $T_c$.

However, when the system approaches a topological transition point at the blue (or red) line in Fig.~2(e), it is expected that the CI phase becomes less stable. This is illustrated in Fig.~4(b) where we show $n_U$ versus $T$ for $\tilde{A} = 0.06$ \AA$^{-1}$. We have used two values of the mass, $m = 3.2$ and $m = 4.4$ meV, which correspond to points just below and just above the topological transition at $m = | \eta | k_D \simeq 3.8$ meV. We notice that for $m = 3.2$ meV, $n_U$ starts decreasing from unity much sooner ($T_c \simeq 8$ K); the gap is $\simeq 2$ meV in this case. On the other hand, for $m = 4.4$ meV, the system is a BI (i.e., it is outside the topological region in the $( m, \tilde{A})$ phase diagram) for which $n_U = 0$ at $T = 0$. For finite $T$, $n_U$ becomes nonzero, which is attributed to thermal activation and partial filling of the conduction band. In the limit $T \rightarrow \infty$, the conduction band has equal population with the valence band, but the curvatures have opposite sign yielding $n_U \rightarrow 0$.
\begin{figure}[t]
\vspace*{-0.0cm}
\begin{center}
\hspace{-0.2cm}\includegraphics[height=4.0cm, width=8.6cm ]{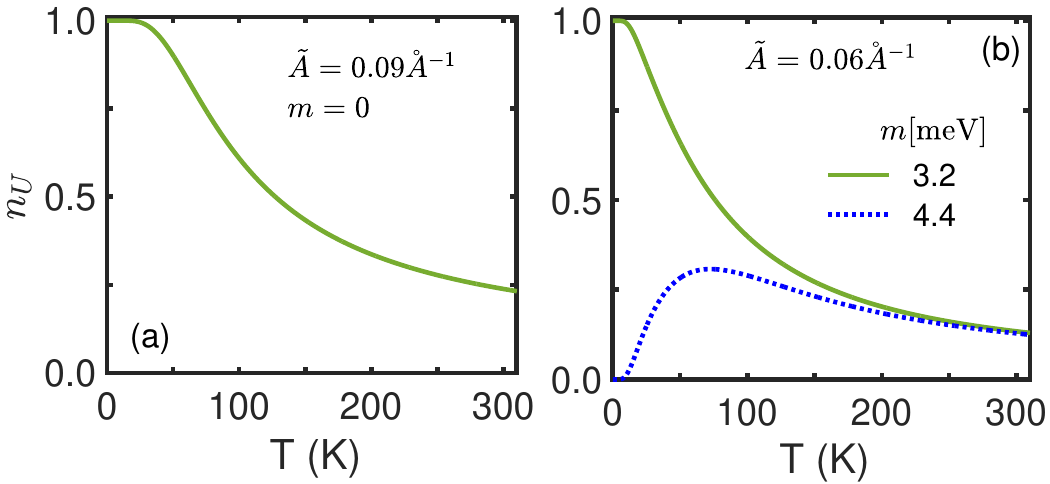}
\end{center}
\vspace*{-0.4cm} \caption{(Colour online) (a) Thermal Uhlmann number $n_U$ as a function of temperature $T$. The system is a CI with $n_U = 1$ for $T < T_c$ with $T_c \simeq 30$ K. For these values of $m$ and $\tilde{A}$ the system is away from the phase boundaries, see Fig.~2(e). (b) The same as in (a) but the values of $m$ are such that the system is just below ($m = 3.2$ meV) or just above ($m = 4.4$ meV) the phase transition.    }
\label{fig:fig1}%
\end{figure}

\section{Orbital magnetization}

The TRS breaking light can generate finite OM, a quantity closely related to anomalous thermoelectric transport. In fact it is the Berry phase correction to the OM that leads to the ANE \cite{niu10}. The OM has been investigated in various systems, such as transition metal dichalcogenides \cite{tahir14,vargiam20}, topological insulators \cite{tahir15}, and Weyl semimetals \cite{tremb19}, to name a few. We obtain the OM from the free energy $F$ in a weak magnetic field $\mathbf{B}$ written as
\begin{equation}
F = - \frac{1}{\beta} \sum_{\lambda \mathbf{k}} \ln \left[ 1 + e^{- \beta ( E_M - E_F )} \right]  ,
\label{eq16}%
\end{equation}
where the electron energy $E_M = E_{\lambda \mathbf{k}} - \boldsymbol{m} ( \mathbf{k} ) \cdot \mathbf{B}$ includes a correction due to the orbital magnetic moment $\boldsymbol{m} ( \mathbf{k} )$. The OM of a band is given by $M_\lambda = - \left( \partial F / \partial B \right)_{E_F,T} = M_{\lambda}^{(o)} + M_\lambda^{(b)}$ where
\begin{equation}
M_{\lambda}^{(o)} = \frac{a^2}{( 2 \pi )^2} \int d^2 k f_{\lambda \mathbf{k} } m_{\lambda} ( \mathbf{k} )   ,
\label{eq17}%
\end{equation}
\begin{equation}
M_{\lambda}^{(b)} = \frac{e a^2}{2 \pi \beta h} \int d^2 k \Omega_{\lambda} ( \mathbf{k} ) \ln \left[ 1 + e^{- \beta ( E_{\lambda \mathbf{k}} - E_F)} \right]    .
\label{eq18}%
\end{equation}
where $a=3.3$ {\AA} is the lattice constant \cite{kim15}. $M_{\lambda}^{(o)}$ is due to the thermodynamic average of the magnetic moments whereas $M_{\lambda}^{(b)}$ results from the Berry phase correction to the electron density of states \cite{niu10}. The OM is given as $M = \sum_{\lambda} M_{\lambda}$. For a two-band model with particle-hole symmetry the orbital magnetic moment and Berry curvature are related by $\boldsymbol{m}_{\lambda} ( \mathbf{k} ) = \left( e / 2h \right) \left( E_\lambda - E_{\lambda^{\prime}} \right) \mathbf{\Omega}_{\lambda} ( \mathbf{k} )$ \cite{niu09}. Using this relation, the OM at zero temperature can be simplified to
\begin{equation}
M_{\lambda} = (1 / e) E_F \sigma_{xy}     .
\label{eq19}%
\end{equation}
This relation is understood by the fact that the Hall conductivity is equal to the derivative of the OM with respect to the chemical potential $E_F$ \cite{nomura17}. When the chemical potential varies in the insulating gap, the OM changes linearly if the Chern number is nonzero; it remains constant otherwise. 
\begin{figure}[t]
\vspace*{0.0cm}
\begin{center}
\hspace{-0.2cm}\includegraphics[height=7.3cm, width=8.6cm ]{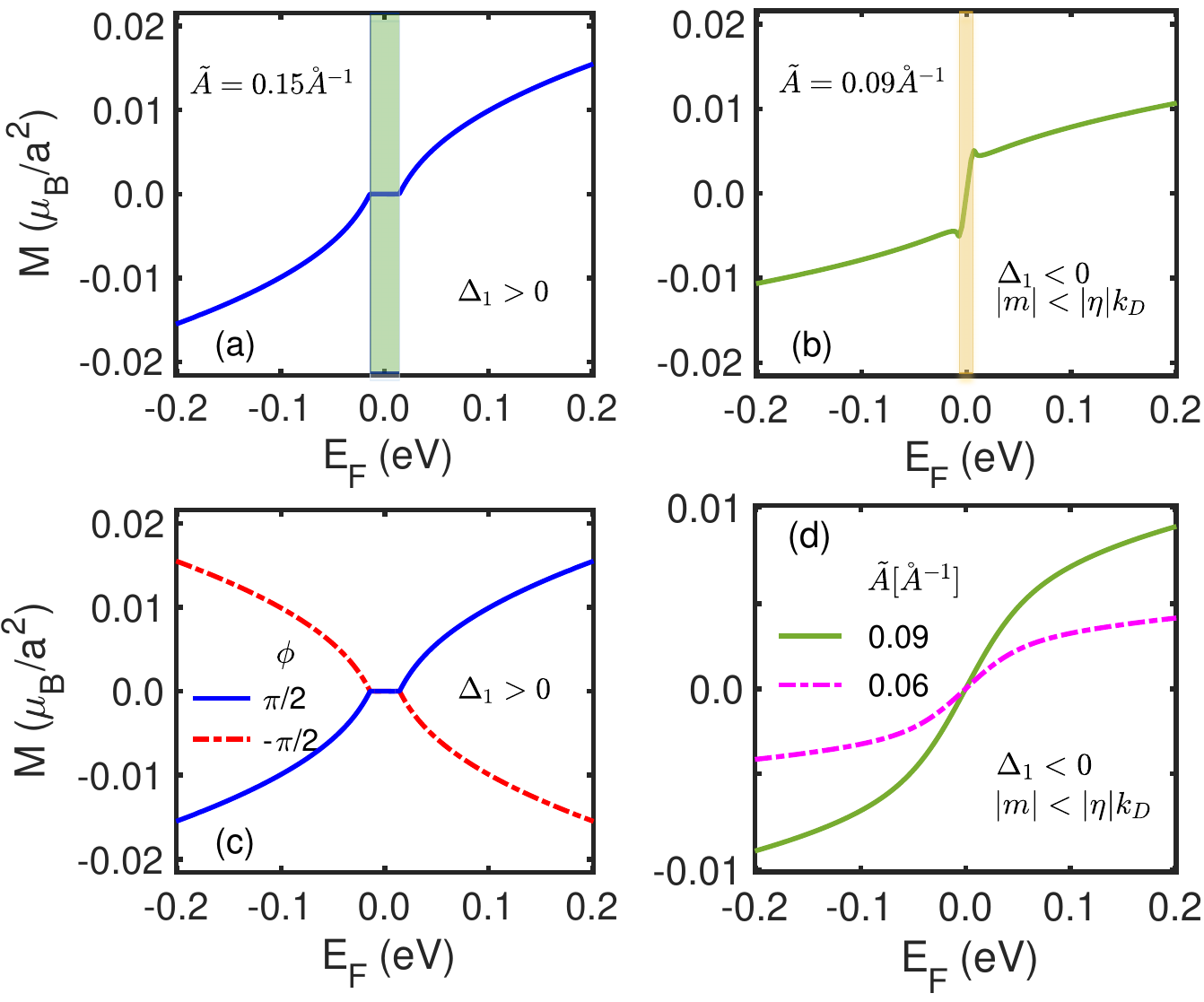}
\end{center}
\vspace*{-0.4cm} \caption{(Colour online) Orbital magnetization $M$ as a function of the chemical potential $E_F$ for an irradiated SD system in the (a) BI and (b) CI phase for $T = 0.1$ K. (c) $M$ as a function of $E_F$ in the BI phase for left- and right-circularly polarized light shown by the blue (solid) and red (dashed-dotted) lines, respectively. (d) $M$ as a function of $E_F$ in the CI phase for $T = 300$ K.   }
\label{fig:fig1}%
\end{figure} 

In Figs.~5(a) and 5(b) we show the OM $M$ versus the chemical potential $E_F$ in the BI and CI phase, respectively, for $\phi = \pi / 2$ and $T=0.1$ K, in units of $\mu_B / a^2$ where $\mu_B = e \hbar / 2 m^\star$ is the Bohr magneton. In this and the next section BI phase corresponds to $\Delta_1 > 0$ and CI to $\Delta_1 < 0$ with $| m | < | \eta | k_D$. We have used $\Delta = - 15$ meV, $\omega = 3 \times 10^{15}$ Hz, and $m = 2$ meV. The values $\tilde{A} = 0.15$ \AA$^{-1}$ and $0.09$ \AA$^{-1}$ ensure that the SD system is a BI and CI, respectively. The shaded areas show the band gaps in the two cases. In the CI phase, the SD system carries a dissipationless chiral edge state that contributes to the OM. The linearity of $M$ on $E_F$ can be seen as the effect of populating the edge state. In the BI phase, we notice that the OM is constant in the gap in accordance with Eq.~(\ref{eq19}).

In Fig.~5(c) we show $M$ versus $E_F$ in the BI phase for left- and right-circularly polarized light, for the same parameter values as in Fig.~5(a). Notice that switching the light polarization induces reversal of $M$. In Fig.~5(d) we show $M$ versus $E_F$ in the CI phase for $T = 300$ K and for two values of $\tilde{A}$. We notice that the linear behaviour of $M$ in the gap is smoothed out due to the Fermi distributions. We also note that the values of $M$ become smaller as the light intensity decreases.

{\section{Charge and thermoelectric transport}

In the presence of an electric field $\mathbf{E}$ and a temperature gradient $\nabla T$, the charge current $\mathbf{J}$ and heat current $\mathbf{J}_Q$ are determined through the following linear response equations:
\begin{equation}
\left(
\begin{array}
[c]{cc}%
\mathbf{J} \\ \
\mathbf{J}_Q
\end{array}
\right) = \left(
\begin{array}
[c]{cc}%
\sigma 
 &  \alpha \\ \
T\alpha  &  \kappa
\end{array}
\right) \left(
\begin{array}
[c]{cc}%
\mathbf{E} \\ \
- \nabla T
\end{array}
\right), \label{eq2000}%
\end{equation}
where $\sigma$, $\alpha$, and $\kappa$ are the charge, thermoelectric, and thermal conductivity tensors, respectively. The off-diagonal (anomalous) components $\sigma_{xy}$ have been evaluated and discussed in the context of the topological phase transition in Sec.~III. In this section we evaluate the diagonal components $\sigma_{aa}$, $\alpha_{aa}$, and $\kappa_{aa}$ using the semiclassical Boltzmann treatment, as well as the off-diagonal components $\alpha_{ab}$ and $\kappa_{ab}$. We start with the latter two which are known as anomalous Nernst and anomalous thermal Hall conductivities; they are topological thermoelectric responses because they arise from a finite Berry curvature.

\subsection{Anomalous Nernst effect}

The conventional Nernst effect \cite{berak13} occurs in the presence of a longitudinal temperature gradient and an external magnetic field, which generates a transverse current $J_i = \alpha_{i j} ( - \nabla_j T )$. A nontrivial Berry curvature also gives rise to a Nernst response, as a result of the anomalous velocity, leading to the ANE. It has been studied in various 2D and Dirac materials \cite{zhou24,vargiam20,berak13,jauho15,sharma18,zhang24}. The ANE is the thermoelectric counterpart of the AHE.

The anomalous Nernst response to a temperature gradient can be obtained using the semiclassical wave packet method together with the contribution of the orbital magnetization \cite{niu10}. In this approach the anomalous Nernst conductivity (ANC) $\alpha_{xy}$ is given as
\begin{equation}
\alpha_{xy} = \frac{e k_B}{\hbar} \sum_{\lambda} \int \frac{d^2 k}{( 2 \pi )^2} \Omega_{\lambda} ( \mathbf{k} ) S_{\lambda}     ,
\label{eq20}%
\end{equation}
where $S_{\lambda} ( \mathbf{k} ) = - f_{\lambda \mathbf{k}} \ln f_{\lambda \mathbf{k}} - ( 1 - f_{\lambda \mathbf{k}} ) \ln (1-f_{\lambda \mathbf{k}})$ is the entropy density. $S_{\lambda} ( \mathbf{k} )$ vanishes for completely filled and completely empty bands, which implies that $\alpha_{xy}$ is a Fermi surface property. At low temperatures the ANC is related to the zero temperature Hall conductivity $\sigma_{xy}$ through the Mott relation \cite{mott69}
\begin{equation}
\alpha_{xy} = - \frac{\pi^2 k_B^2 T}{3 e} \frac{d \sigma_{xy}}{d E_F}    .
\label{eq21}%
\end{equation}

The ANC versus the chemical potential is plotted in Figs.~6(a) and 6(b) for the BI and CI phase, respectively, for $\phi = \pi / 2$ and $T = 10$ K. We have used the same parameter values as in Fig.~5. In Fig.~6(a) we notice that $\alpha_{xy} = 0$ when $E_F$ is in the band gap whereas at the band edges it exhibits a step jump. This can be understood from the Mott relation; for $E_F$ in the gap, $\sigma_{xy}$ is constant (see Fig.~3) and thus $d \sigma_{xy} / d E_F = 0$. At the band edges $d \sigma_{xy} / d E_F$ is discontinuous resulting in the sharp dip or peak in $\alpha_{xy}$. Their opposite signs reflect those of the Berry curvatures of the respective bands. As $E_F$ shifts away from the band edges, $\alpha_{xy}$ decays rapidly and eventually goes to zero; the magnitude of the Berry curvature decreases for electron or hole states away from the SD nodes. We verified numerically that the decay follows $\sim 1 / E_F^2$.
\begin{figure}[t]
\vspace*{0.0cm}
\begin{center}
\hspace{-0.2cm}\includegraphics[height=7.6cm, width=8.6cm ]{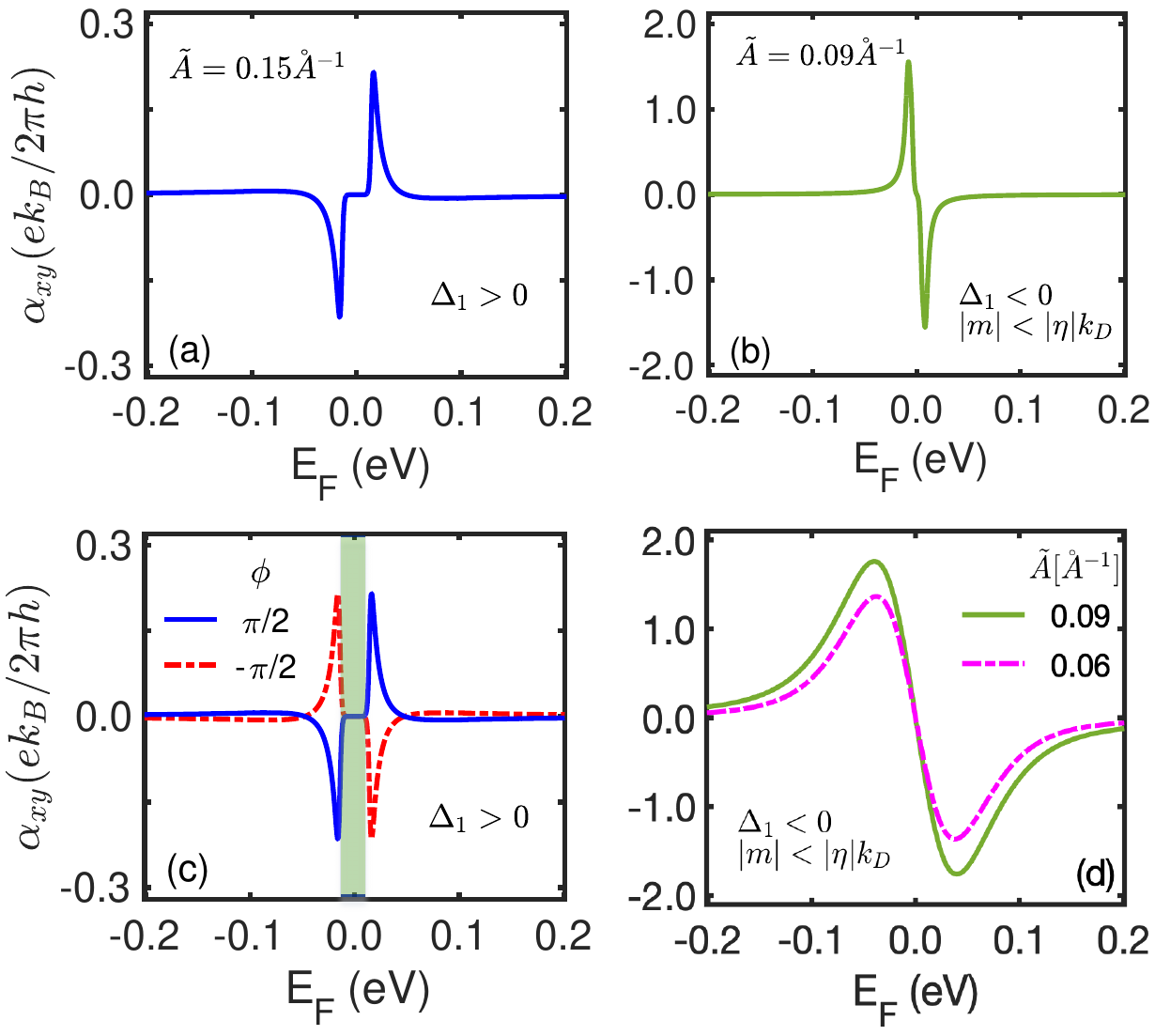}
\end{center}
\vspace*{-0.4cm} \caption{(Colour online) ANC $\alpha_{xy}$ versus $E_F$ for an irradiated SD system in the (a) BI and (b) CI phase for $T = 10$ K. (c) $\alpha_{xy}$ versus $E_F$ in the BI phase for left- and right-circularly polarized light shown by the blue (solid) and red (dashed-dotted) lines, respectively. (d) $\alpha_{xy}$ versus $E_F$ in the CI phase for $T = 300$ K.   }
\label{fig:fig1}%
\end{figure} 

In Fig.~6(b) $\alpha_{xy}$ exhibits approximately a $10$-fold increase. This is a consequence of the value of the light intensity; it shrinks the band gap and brings the system close to a SSDC state resulting in significantly enhanced Berry curvature. We also notice that the dip-peak profile is reversed compared to the BI phase; it is a consequence of the sign reversal in the Berry curvature for $\Delta_1 < 0$.

In Fig.~6(c) we show the ANC in the BI phase for left- and right-circularly polarized light. The parameter values are the same as in Fig.~6(a). We notice that switching the polarization of light induces reversal of $\alpha_{xy}$. This effect can also be observed in the CI phase. Thus the light polarization can be used to reverse the direction of transverse flow of thermoelectric charge current in a SD material. The effect of temperature is shown in Fig.~6(d) where we plot $\alpha_{xy}$ versus $E_F$ for $T = 300$ K. We notice that the sharp jumps shown in Fig.~6(b) are smoothed out at high temperatures. At the peak we estimate $\alpha_{xy} \simeq 1$ nA/K, which is an order of magnitude larger than that of transition metal dichalcogenides \cite{vargiam20,jauho15,sharma18} and of strained graphene \cite{berak13}, partly because of the larger values of the Berry curvature. However it is comparable to that of silicene \cite{jin16} and jacutingaite \cite{zhang24}.  

\subsection{Anomalous Thermal Hall effect}

The thermal Hall effect occurs in the presence of a temperature gradient in the longitudinal direction and an out-of-plane magnetic field, which generates a heat current in the transverse direction, $J_{Q,i} = \kappa_{ij} ( - \nabla_j T )$ \cite{gold}. However, as in the Nernst effect, a nontrivial Berry curvature can play the role of a magnetic field, leading to the anomalous thermal Hall effect (ATHE).

The anomalous thermal Hall conductivity (ATHC) $\kappa_{xy}$ is given by 
\begin{eqnarray}
\nonumber \hspace*{-0.2cm} \kappa_{xy}  &=& \frac{k_B^2 T}{4 \pi^2 \hbar}  \sum_{\lambda} \int d^2 k
\Omega_{\lambda} ( \mathbf{k} ) \Big\{ \frac{\pi^2}{3} + \beta^2 ( E - E_F )^2 f_{\lambda \mathbf{k}} \\* 
&&-2 \text{Li}_{2} ( 1 - f_{\lambda \mathbf{k}} ) - \left[ \ln ( 1 - f_{\lambda \mathbf{k}} ) \right]^2 \Big\}  ,
\label{eq22}%
\end{eqnarray}
where $\text{Li}_{n}( z ) = \sum_{k=1}^{\infty} z^k / k^n$ is the polylogarithmic function. At low temperatures the ATHC is related to the zero temperature Hall conductivity $\sigma_{xy}$ through the Wiedemann-Franz law \cite{mahan} $\kappa_{xy} = \left( \pi^2 k_B^2 T / 3 e^2 \right) \sigma_{xy}$. This relation indicates that $\kappa_{xy}$ exhibits similar beviour to $\sigma_{xy}$; they differ only by a factor.
\begin{figure}[t]
\vspace*{0.0cm}
\begin{center}
\hspace{-0.2cm}\includegraphics[height=7.6cm, width=8.6cm ]{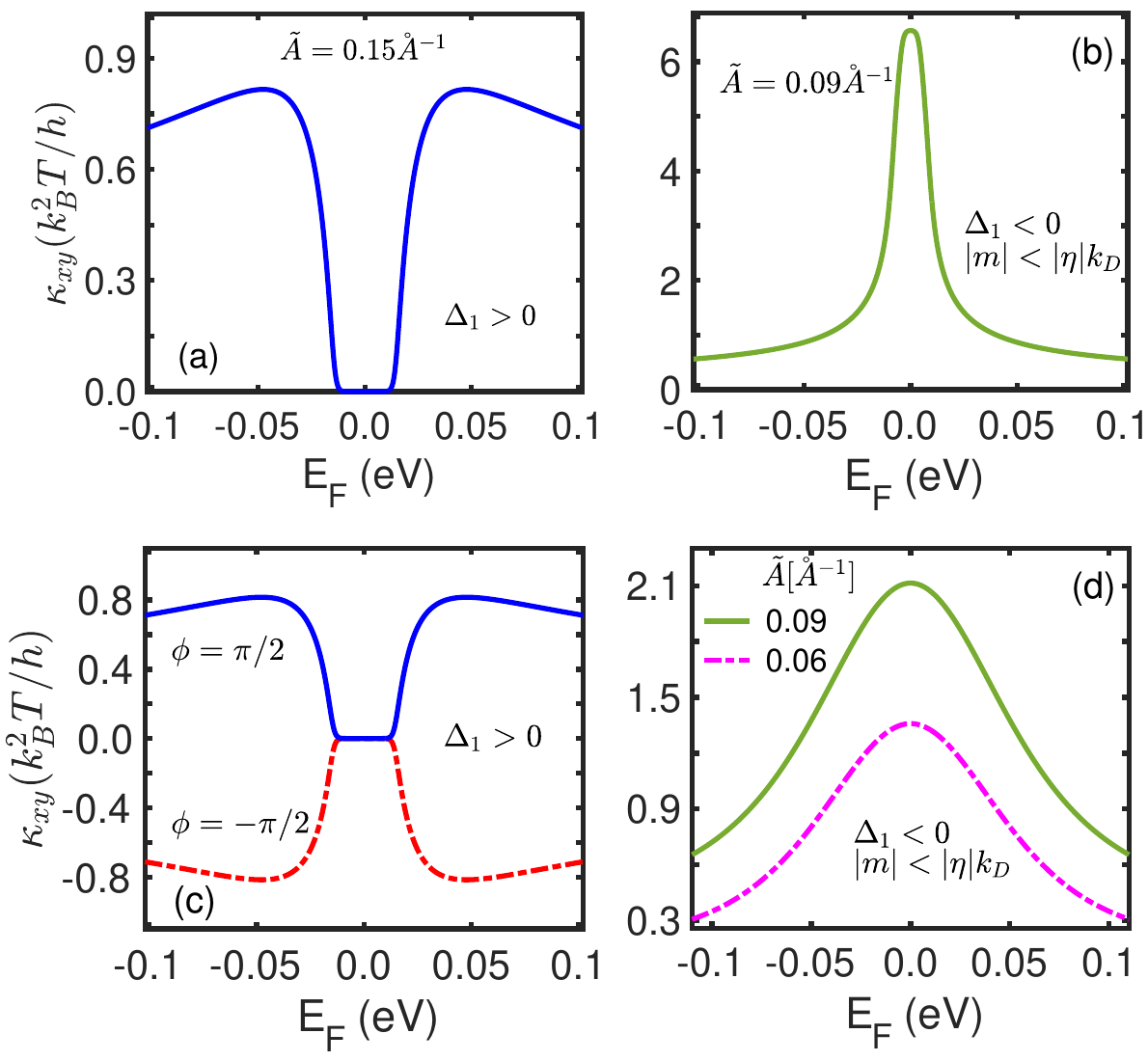}
\end{center}
\vspace*{-0.4cm} \caption{(Colour online) ATHC $\kappa_{xy}$ versus $E_F$ for an irradiated SD system in the (a) BI and (b) CI phase for $T = 10$ K. (c) $\kappa_{xy}$ versus $E_F$ in the BI phase for left- and right-circularly polarized light shown by the blue (solid) and red (dashed-dotted) lines, respectively. (d) $\kappa_{xy}$ versus $E_F$ in the CI phase for $T = 200$ K.   }
\label{fig:fig1}%
\end{figure} 

In Figs.~7(a) and 7(b) we show $\kappa_{xy}$ versus $E_F$ evaluated using Eq.~(\ref{eq22}) for the BI and CI phase, respectively, for $\phi = \pi / 2$ and $T = 10$ K. The parameter values are the same as in Fig.~6. We notice that, in the BI phase the ATHC vanishes when $E_F$ is inside the gap. In the CI phase, $\kappa_{xy}$ increases drastically and exhibits a much narrower plateau. In this case, the system is near a SSDC state, i.e., near a band crossing, which generates enhanced Berry curvature which then gives a large contribution to $\kappa_{xy}$.

In Fig.~7(c) we show the ATHC in the BI phase for left- and right-circular polarizations. As in the ANE, we notice that switching the polarization of the incident light from left- to right-circular induces reversal of $\kappa_{xy}$. Thus the consequence of switching the polarization manifests as a reversal of heat flow. Such a reversal of heat flow could be used to exploit the phenomenon of thermal rectification \cite{rectification}.

In Fig.~7(d) we plot $\kappa_{xy}$ versus $E_F$ for $T = 200$ K. We notice the thermal broadening of the ATHC due to the Fermi distributions. Note also that $\kappa_{xy}$ is suppressed for lower light intensity.

\subsection{Robustness of the topological responses}

The results for the topological responses $\sigma_{xy}$, $\alpha_{xy}$, $\kappa_{xy}$, and $M$ presented above in the BI and CI phases depend strictly on the parameter $\Delta_1 = \Delta + \alpha \tilde{A}^2 / 2$ and the value of the mass $m$. Focusing on $\Delta_1$ first, these responses exhibit a robust pattern of behaviour in the BI and CI phases as long as $\Delta_1 > 0$ and $\Delta_1 < 0$ (with $| m | < | \eta | k_D$), respectively. However, varying either $\Delta$ or $\tilde{A}$ or both in a way that make $\Delta_1$ smaller, leads to larger magnitudes of $\alpha_{xy}$ and $\kappa_{xy}$. This is due to the fact that the band gap depends on $\Delta_1$; a smaller $\Delta_1$ leads to smaller band gap, which generates larger Berry curvature. Consequently $\alpha_{xy}$ and $\kappa_{xy}$ are enhanced. We remark that for $\Delta = 0$, the parameter $\Delta_1 = \alpha \tilde{A}^2 / 2 > 0$ and the SD system is in the BI phase with $\mathcal{C} = 0$. Focusing now on the value of $| m |$ (with $\Delta_1 < 0$), the topological responses are enhanced when $| m |$ approaches a SSDC gapless state, see Fig.~2(e). As $| m |$ increases for fixed $\tilde{A}$ and approaches the phase boundaries, $| m | \rightarrow | \eta | k_D$, the band gap reduces and $\Omega ( \mathbf{k} )$ increases leading to larger $\alpha_{xy}$ and $\kappa_{xy}$.

\subsection{Charge conductivities $\sigma_{xx}$ and $\sigma_{yy}$}

We start with the evaluation of the density of states (DOS) and then proceed with the charge conductivities $\sigma_{xx}$ and $\sigma_{yy}$. In this and the following sections we set $m = 0$.

\textit{Density of states}. In order to avoid difficulties associated with the anisotropic energy dispersion, we transform coordinates from $(k_x, k_y)$ to new variables $(q_x, q_y)$ defined by $q_x = \Delta_1 + \alpha k_x^2$ and $q_y = \gamma k_y$. The Jacobian of this transformation is given by \cite{range}
\begin{equation}
J ( q, \theta ) = \frac{1}{2 \gamma \sqrt{\alpha}} \left( q \cos \theta - \Delta_1 \right)^{-1/2}    ,
\label{eq23}%
\end{equation}
where $\tan \theta = q_y / q_x$ and $q = ( q_x^2 + q_y^2 )^{1/2}$. The DOS $D(E)$ can then be obtained from
\begin{equation}
D ( E ) = \frac{1}{( 2 \pi )^2} \int d^2 q J( q, \theta ) \delta ( E_{\lambda q} - E )    ,
\label{eq24}%
\end{equation}
where $E_{\lambda q} = \lambda \left[ q^2 + ( \eta^2 / \alpha) ( q \cos \theta - \Delta_1 )  \right]^{1/2}$. We have evaluated the integral both numerically and analytically. In the analytical evaluation we made the approximation $- \nu \cos \theta + (E^2 + \mu)^{1/2} \simeq (E^2 + \mu)^{1/2}$, where $\nu = \eta^2 / 2 \alpha$ and $\mu = \eta^2 \Delta_1 / \alpha$, which is valid for low light intensities. We find that the results agree extremely well with the numerical ones. Small deviations occur in the case $\Delta_1 < 0$ for very low energies. For $\Delta_1 < 0$ we find
\begin{equation}
D ( E ) = \frac{D_0 | E |}{\sqrt{\epsilon - \Delta_1}}  K ( r )  ,  \hspace{0.25in}   \sqrt{| \mu |} < | E | <  | \Delta_1 | ,
\label{eq25}%
\end{equation}
\begin{equation}
D ( E ) = \frac{D_0 | E |}{\sqrt{ 2 \epsilon}}  K \left( 1 / r \right)  ,  \hspace{0.69in}  | E | > | \Delta_1 | ,
\label{eq26}%
\end{equation}
where $K(x)$ is the complete elliptic integral of the first kind \cite{table}, $\epsilon = ( E^2 + \mu )^{1/2}$, $D_0 = \sqrt{2 m} / 2 \pi^2 \hbar^2 v_y$, and $r = [ 2 \epsilon / ( \epsilon - \Delta_1)]^{1/2}$. At the SD transition point, $\Delta_1 = 0$, we find
\begin{equation}
D ( E ) = \frac{D_0 | E |}{\sqrt{2 \epsilon}} K ( 1 / \sqrt{2} )  ,
\label{eq27}%
\end{equation}
where $K(1/\sqrt{2}) = 1.85$. Above the transition point, $\Delta_1 > 0$, we find
\begin{equation}
D ( E ) = 0  , \hspace{1in}  | E | <  \Delta_1 ,
\label{eq28}%
\end{equation}
\begin{equation}
D ( E ) = \frac{D_0 | E | }{\sqrt{2 \epsilon}} K \left( 1 / r \right)  , \hspace{0.22in}  | E | >  \Delta_1 .
\label{eq29}%
\end{equation}

In Fig.~8 we show $D ( E )$ in units of $D_0$ for an irradiated SD system in the semimetallic phase with two gapped Dirac nodes (green line,  $\Delta_1 < 0$), in the SD transition point (red line, $\Delta_1 = 0$), and in the BI phase (blue line, $\Delta_1 > 0$). Here we have used $\omega = 5 \times 10^{15}$ Hz, $\Delta = - 30$ meV, and a left-circularly polarized light ($\phi = \pi / 2$). For $\Delta_1 < 0$ we notice that $D(E)$ exhibits van Hove singularities at $| E | = | \Delta_1 |$ due to saddle points of the energy dispersion; for non-irradiated SD systems \cite{fuch09} they occur at $| E | = | \Delta |$. For $\Delta_1 = 0$, $D(E)$ is $\sim \sqrt{E}$. For $\Delta_1 > 0$, $D(E)$ exhibits a jump at $| E | = \Delta_1$ and increases $\sim \sqrt{E}$ thereafter. We also show the approximate analytical result for the case $\Delta_1 = 0$ by the black dashed line. For the other cases (not shown here) we also find very good agreement with the numerical results.
\begin{figure}[t]
\vspace*{0.0cm}
\begin{center}
\hspace{-0.2cm}\includegraphics[height=5.6cm, width=7.6cm ]{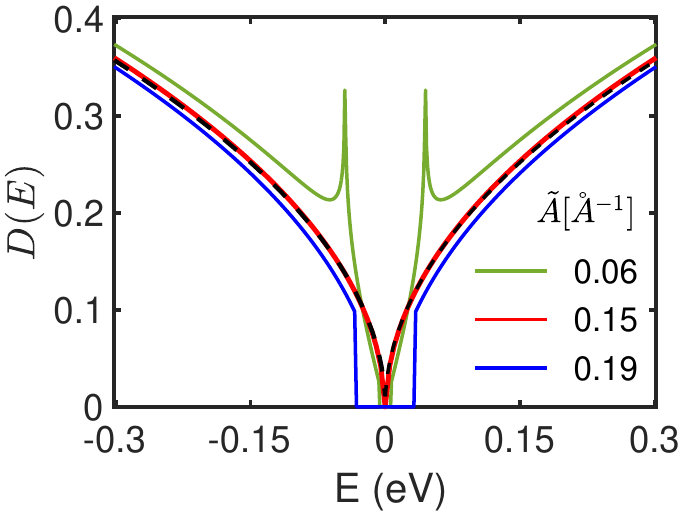}
\end{center}
\vspace*{-0.4cm} \caption{(Colour online) Density of states $D(E)$ in units of $D_0$ for increasing values of the light intensity $\tilde{A}$. For $\tilde{A} = 0.06$ \AA$^{-1}$ ($\Delta_1 < 0$) $D(E)$ exhibits van Hove singularities, whereas for $\tilde{A} = 0.19$ \AA$^{-1}$ ($\Delta_1 > 0$) it exhibits  jump discontinuities. At the SD transition point ($\tilde{A} = 0.15$ \AA$^{-1}$, $\Delta_1 = 0$) $D(E)$ varies as $\sqrt{E}$. The black dashed line is the approximate analytical result Eq.~(\ref{eq27}), for $\tilde{A} = 0.15$ \AA$^{-1}$.   }
\label{fig:fig1}%
\end{figure} 

\textit{Charge conductivities}. Using the Boltzmann transport theory in the relaxation-time approximation we obtain the diagonal components of the conductivity tensor as
\begin{equation}
\sigma_{aa} = e^2 \sum_{\lambda = \pm} \int \frac{d^2 k}{(2 \pi )^2} v_{a, \lambda \mathbf{k}}^2 \tau_{\lambda \mathbf{k}} \left( - \frac{\partial f_{\lambda \mathbf{k}}}{\partial E_{\lambda \mathbf{k}}} \right)  ,
\label{eq30}%
\end{equation}
where $a = x, y$, $\tau_{\lambda \mathbf{k}}$ is the relaxation time, and $v_{a, \lambda \mathbf{k}} = \hbar^{-1} \partial_{k_{a}} E_{\lambda \mathbf{k}}$ the band velocities. For simplicity we will treat $\tau_{\lambda \mathbf{k}}$ as independent of momentum, $\tau_{\lambda \mathbf{k}} = \tau_F$. The band velocities are easily found to be
\begin{equation}
v_{x, \lambda \mathbf{k}} = \lambda \hbar^{-1} \varepsilon_{\mathbf{k}}^{-1} ( 2 \alpha^2 k_x^3 + \Lambda k_x )   ,
\label{eq31}%
\end{equation}
\begin{equation}
v_{y, \lambda \mathbf{k}} = \lambda \hbar^{-1} \varepsilon_{\mathbf{k}}^{-1} \gamma^2 k_y    ,
\label{eq32}%
\end{equation}
where $\Lambda = 2 \alpha \Delta_1 + \eta^2$, and $\varepsilon_{\mathbf{k}}$ is given in Eq.~(\ref{eq9}) with $m = 0$. We have evaluated numerically the above integrals for zero and finite temperatures (see Fig.~9). However, using the approximation $- \partial f_{\lambda \mathbf{k}} / \partial E_{\lambda \mathbf{k}} \simeq \delta ( E_{\lambda \mathbf{k}} - E_F )$, valid for very low temperatures, we obtain analytic expressions for the conductivities when $\Delta_1 = 0$. Transforming $( k_x, k_y )$ to the variables $(q, \theta)$ and taking $E_F$ in the electron band we obtain
\begin{equation}
\sigma_{xx} = \frac{e^2}{h} \frac{\tau_F}{\pi \hbar \gamma \sqrt{\alpha}} \frac{1}{E_F} ( I_1 + I_2 )    ,
\label{eq33}%
\end{equation}
where
\begin{equation}
I_1 = \alpha \int ( \epsilon_F \cos \theta - \Delta_1 )^{5/2} d \theta  ,
\label{eq34}%
\end{equation}
\begin{equation}
I_2 = \Lambda \int ( \epsilon_F \cos \theta - \Delta_1 )^{3/2} d \theta   ,
\label{eq35}%
\end{equation}
with $\epsilon_F = ( E_F^2 + \mu )^{1/2}$. In obtaining Eq.~(\ref{eq33}) we made the same approximations as for the DOS. We have also ignored terms $\sim \Lambda^2 \sim \eta^4 / \alpha^2$, which are negligible for low light intensities. For $\sigma_{yy}$ we obtain
\begin{equation}
\sigma_{yy} = \frac{e^2}{h} \frac{\gamma \tau_F}{4 \pi \hbar \sqrt{\alpha}} \frac{\epsilon_F^2}{E_F} I_3   ,
\label{eq36}%
\end{equation}
where
\begin{equation}
I_3 = \int \sin^2 \theta ( \epsilon_F \cos \theta - \Delta_1 )^{-1/2} d \theta   .
\label{eq37}%
\end{equation}

For the SD transition point, $\Delta_1 = 0$, the integrals $I_1$, $I_2$, and $I_3$ can be computed analytically; the range of $\theta$ in this case is $-\pi/2 < \theta < \pi/2$. They are given by
\begin{equation}
I_1 = 2 \alpha E_F^{5/2} \int_0^1 x^{5/2} ( 1 - x^2 )^{-1/2} dx = 2 \alpha c_1 E_F^{5/2}    ,
\label{eq38}%
\end{equation}
\begin{equation}
I_2 = 2 \Lambda E_F^{3/2} \int_0^1 x^{3/2} ( 1 - x^2 )^{-1/2} dx = 2 \Lambda c_2 E_F^{3/2}    ,
\label{eq39}%
\end{equation}
\begin{equation}
I_3 = 2 E_F^{-1/2} \int_0^1 x^{-1/2} ( 1 - x^2 )^{1/2} dx = c_3 E_F^{-1/2}    ,
\label{eq40}%
\end{equation}
where $c_1 = 0.7189$, $c_2 = 0.874$, and $c_3 = 3.496$. With the help of Eqs.~(\ref{eq38}) - (\ref{eq40}) we can finally express the charge conductivities as
\begin{equation}
\sigma_{xx} = \frac{e^2}{h} \frac{2 \tau_F}{\pi \hbar \gamma \sqrt{\alpha}} \left( \alpha c_1 E_F^{3/2} + \eta^2 c_2 E_F^{1/2} \right)    ,
\label{eq41}%
\end{equation}
\begin{equation}
\sigma_{yy} = \frac{e^2}{h} \frac{\gamma \tau_F}{4 \pi \hbar \sqrt{\alpha}} c_3 E_F^{1/2}   ,
\label{eq42}%
\end{equation}
We notice that, in the absence of light, $\eta = 0$, the power-law dependence of $\sigma_{xx}$ on $E_F$ is given by $\sigma_{xx} \sim E_F^{3/2}$. The light manifests in the second term of Eq.~(\ref{eq41}). On the other hand, $\sigma_{yy} \sim E_F^{1/2}$ independent of the light.

In Figs.~9(a) and 9(c) we show numerical results for $\sigma_{xx}$ and $\sigma_{yy}$, respectively, as functions of $E_F$ for $T = 0$ K and the same parameter values as in Fig.~8. We have used $\tau_F = 160$ fs, which is relevant for a monolayer black phosphorus \cite{time}. We notice that the conductivities exhibit distinct dependence on $E_F$. In the gapped semimetallic phase (green solid line, $\Delta_1 < 0$) we notice a change in slope for both $\sigma_{xx}$ and $\sigma_{yy}$ near the van Hove singularity, i.e., for $E_F \simeq ( | \mu | + \Delta_1^2 )^{1/2} \simeq 0.04$ eV (see Fig.~8). For $E_F < ( | \mu | + \Delta_1^2 )^{1/2}$, both conductivities increase linearly but the slope of $\sigma_{yy}$ is approximately ten times greater; this is a manifestation of the linear energy dispersion in $k_y$. In fact, we notice from Eqs.~(\ref{eq41}) and (\ref{eq42}) that, in the absence of light, $\sigma_{xx} / \sigma_{yy} = ( 4 c_1 / m v_y^2 c_3 ) E_F \sim E_F$, which can be very small, especially for low doping levels. This is a consequence of the much larger square velocity $v_y^2$ in the direction with the linear dispersion compared to $v_x^2$. In the absence of light, Eqs.~(\ref{eq31}) and (\ref{eq32}) become $v_x^2 = 4 \alpha E_F / \hbar^2$ (for $k_y = 0$) and $v_y^2 = \gamma^2 / \hbar^2=$ constant (for $k_x = 0$) yielding $v_x^2 / v_y^2 = (2 / m v_y^2) E_F \sim E_F$ \cite{effective}.
\begin{figure}[t]
\vspace*{0.0cm}
\begin{center}
\hspace{-0.2cm}\includegraphics[height=7.1cm, width=8.6cm ]{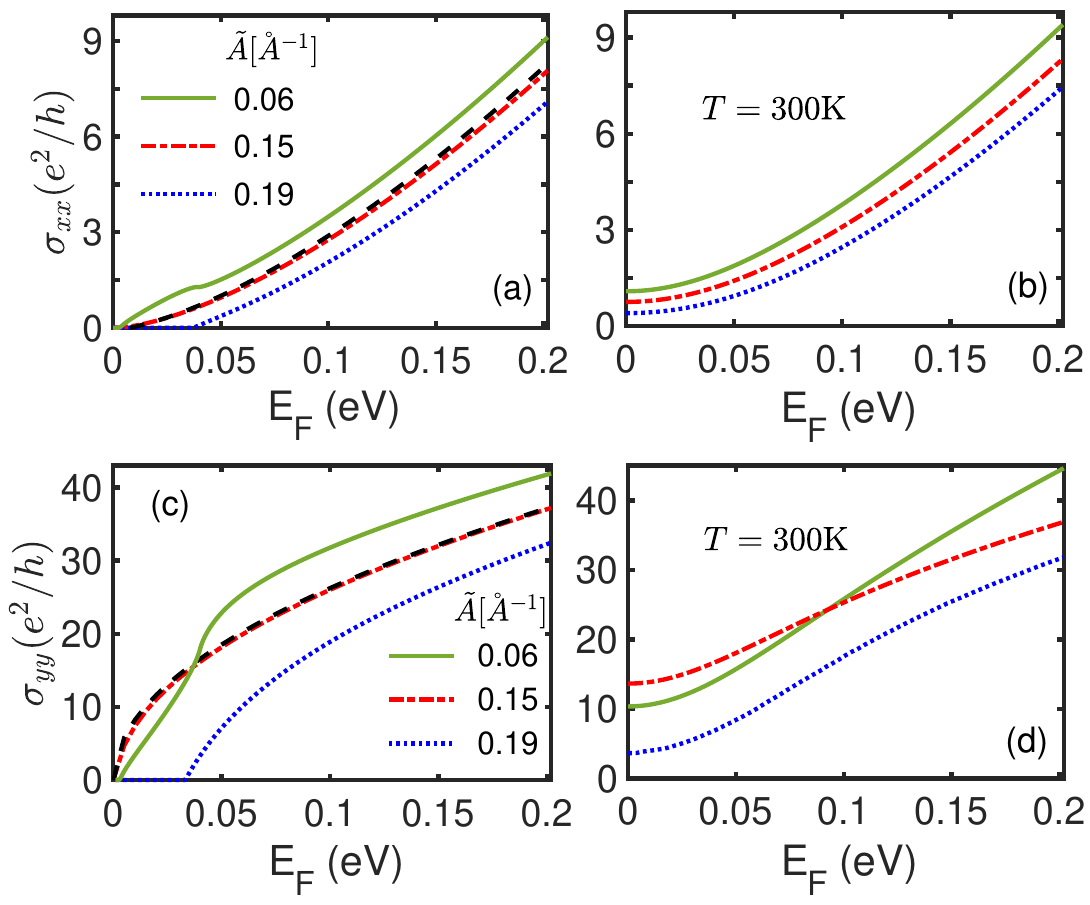}
\end{center}
\vspace*{-0.4cm} \caption{(Colour online) (a) Charge conductivity $\sigma_{xx}$ versus $E_F$  for increasing values of the light intensity $\tilde{A}$ and $T = 0$ K. The black dashed line is the analytical result Eq.~(\ref{eq41}) for the SD transition point. (b) The same as in (a) but for $T = 300$ K. (c) Charge conductivity $\sigma_{yy}$ versus $E_F$ for increasing values of $\tilde{A}$ and $T = 0$ K. The black dashed line is the analytical result Eq.~(\ref{eq42}). (d) The same as in (c) but for $T = 300$ K.   }
\label{fig:fig1}%
\end{figure} 

For the SD point ($\Delta_1 = 0$), besides the numerical results (red dashed-dotted lines) we also plot the analytical results (black dashed lines) from Eqs.~(\ref{eq41}) and (\ref{eq42}), finding that the later is an excellent approximation. In this case, $\sigma_{xx}$ increases superlinearly but $\sigma_{yy} \sim E_F^{1/2}$. In Figs.~9(b) and 9(d) we show $\sigma_{xx}$ and $\sigma_{yy}$, respectively, for $T = 300$ K. We notice that, for low values of $E_F$, the finite temperature raises the values of the conductivities. We also notice that the change in slope for $T = 0$ K is smoothed out due to the thermal broadening. We remark that these results are independent of the polarization of the light because of the occurrence of $\eta^2$ in the band velocities. This is in contrast to the transverse (topological) responses where $\Omega ( \mathbf{k} )$ depends linearly on $\eta$. 

\subsection{Thermoelectric conductivities $\alpha_{xx}$ and $\alpha_{yy}$}

We obtain the diagonal components of the thermoelectric conductivity tensor as
\begin{equation}
\alpha_{aa} = \frac{- e}{T} \sum_{\lambda = \pm} \int \frac{d^2 k}{(2 \pi)^2} v_{a, \lambda \mathbf{k}}^2 \tau_{\lambda \mathbf{k}}  ( E_{\lambda \mathbf{k}} - E_F) \left(  - \frac{\partial f_{\lambda \mathbf{k}}}{\partial E_{\lambda \mathbf{k}}} \right)  ,
\label{eq43}%
\end{equation}
where $\alpha = x, y$. We show numerical results for $\alpha_{xx}$ and $\alpha_{yy}$ versus $E_F$ in Figs.~10(a) and 10(b), respectively, for $T = 300$ K. The parameter values are the same as in Fig.~9. In general, the thermoelectric conductivity tensor $\alpha_{aa}$ is proportional to the negative derivative of the charge conductivity tensor $\sigma_{aa}$ with respect to the chemical potential, $\alpha_{aa} \sim  - \partial \sigma_{aa} / \partial E_F $. In the limit $E_F \rightarrow 0$, we notice that $\alpha_{aa} = 0$; the variation in $\sigma_{aa}$ caused by an infinitesimal shift in $E_F$ vanishes, see Figs.~9(b) and 9(d). As $E_F$ increases $|\alpha_{xx}|$ increases because the variation in $\sigma_{xx}$ becomes nonzero. However, the corresponding rate of increase of $|\alpha_{yy}|$ is at least five times larger than $|\alpha_{xx}|$; the slope of $\sigma_{yy}$ is much greater than that of $\sigma_{xx}$. This leads to values of $|\alpha_{yy}|$ which are an order of magnitude larger than those of $|\alpha_{xx}|$. For higher values of $E_F$, $|\alpha_{yy}|$ tends to a constant value $|\alpha_{yy}| \simeq 100 \hspace{0.1cm} ek_B/2\pi h \simeq 53$ nA/K.

We also notice that the effect of the light on the two thermoelectric responses is different. Even though $\alpha_{xx}$ depends weakly on the light intensity $\tilde{A}$, it is nevertheless suppressed as $\tilde{A}$ increases. Its highest values are attained when the SD system is in the gapped semimetallic phase, $\tilde{A} = 0.06$ \AA$^{-1}$. In contrast, $\alpha_{yy}$ depends strongly on the light intensity and it is enhanced as $\tilde{A}$ increases; its highest values are attained in the BI phase, $\tilde{A} = 0.19$ \AA$^{-1}$. For $E_F = 0.8$ eV and $\tilde{A} = 0.19$ \AA$^{-1}$, we estimate their values to be $\alpha_{xx} \simeq - 9$ nA/K and $\alpha_{yy} \simeq - 97$ nA/K.

\subsection{Thermal conductivities $\kappa_{xx}$ and $\kappa_{yy}$}

We obtain the diagonal components of the thermal conductivity tensor as
\begin{equation}
\kappa_{aa} \hspace{-0.025cm} = \hspace{-0.025cm} \frac{ -1}{T} \sum_{\lambda = \pm} \hspace{-0.025cm} \int \hspace{-0.055cm} \frac{d^2 k}{(2 \pi)^2} \hspace{0.04cm} v_{a, \lambda \mathbf{k}}^2 \tau_{\lambda \mathbf{k}}  ( E_{\lambda \mathbf{k}} - E_F)^2 \hspace{-0.02cm} \left(-  \frac{\partial f_{\lambda \mathbf{k}}}{\partial E_{\lambda \mathbf{k}}} \right)  ,
\label{eq44}%
\end{equation}
where $\alpha = x, y$. In Figs.~11(a) and 11(b) we show the thermal conductivities $\kappa_{xx}$ and $\kappa_{yy}$, respectively, as functions of $E_F$ for $T = 300$ K, and for the same parameter values as in Fig.~9. Their dependence on $E_F$ is similar to those of the charge conductivities. Similarly to the thermoelectric conductivities, we notice that the dependence of $\kappa_{yy}$ on the light intensity $\tilde{A}$ is much stronger than that of $\kappa_{xx}$. We remark that $\kappa_{yy}$ can be two orders of magnitude larger than $\kappa_{xx}$. For $E_F = 0$ and $\tilde{A} = 0.15$ \AA$^{-1}$, we estimate their values to be $\kappa_{xx} \simeq 6.9 \times 10^{-10}$ W/K and $\kappa_{yy} \simeq 2.6 \times 10^{-8}$ W/K.

\section{Possible opto-thermoelectric applications}

The reversal of the direction of flow of the anomalous (topological) currents by flipping the polarization of light is promising for possible opto-thermoelectric applications and devices. For instance, the reversal and control of the ATHC $\kappa_{xy}$ with light suggests a possible pathway for the exploitation of the phenomenon of thermal rectification \cite{rectification}, which is desirable in quantum \cite{gupt22} and magnetic \cite{cast23} thermal transistors. The former are analogous to electronic transistors with temperatures replacing voltages and thermal energy replacing electric current. The latter use a temperature-dependent magnetization of a ferromagnetic material to control the source-drain heat flow by a gate temperature in a manner that leads to heat flow switching. Our results suggest that, heat flow switching could be achieved in irradiated Dirac or semi-Dirac materials by changing the polarization of light, which leads to a sign change of the Berry curvature-the driver of the anomalous topological responses.
\begin{figure}[t]
\vspace*{0.0cm}
\begin{center}
\hspace{-0.2cm}\includegraphics[height=3.4cm, width=8.8cm ]{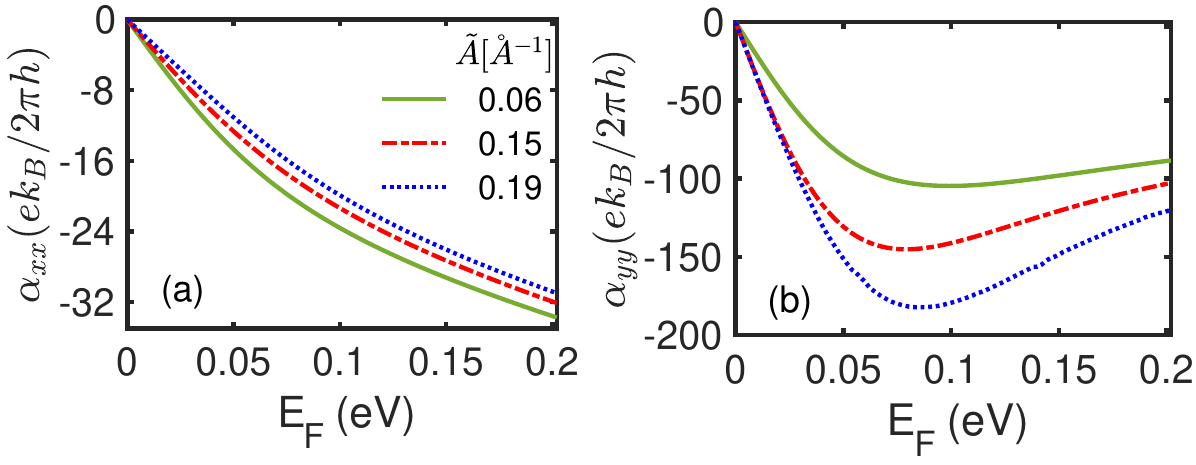}
\end{center}
\vspace*{-0.4cm} \caption{(Colour online) Thermoelectric conductivities (a) $\alpha_{xx}$ and (b) $\alpha_{yy}$ versus $E_F$ for increasing values of the light intensity $\tilde{A}$. The temperature is $T = 300$ K.  }
\label{fig:fig1}%
\end{figure}

The light-induced sign change of the Berry curvature also leads to a sign reversal of the Nernst signal $\mathcal{N}= E_y / \nabla_x T = ( \alpha_{xy} \sigma_{xx} - \alpha_{xx} \sigma_{xy} ) / ( \sigma_{xx}^2 + \sigma_{xy}^2 )$. Switching the polarization of light results to a sign change of $\alpha_{xy}$ and of $\sigma_{xy}$ with $\alpha_{xx}$ and of $\sigma_{xx}$ remaining unchanged. This leads to reversal of the polarity of $\mathcal{N}$, which coud be useful for switchable transverse thermoelectric generation \cite{yama21} without using magnetic materials.

\section{Summary and conclusions}

We investigated the topological phase transitions in a semi-metallic SD material under the application of off-resonant light and with an inversion symmetry breaking mass. We derived analytical expressions for the energy spectrum, the Berry curvature, the Chern numbers, and presented the $(m, \tilde{A})$ phase diagram. We also showed the existence of single semi-Dirac-cone semimetal states. These  remarkable states  emerge along the phase boundaries; a hallmark of them is that one SD cone is gapless and the other  gapped. We showed that, as the light intensity increases for fixed mass, the system undergoes normal-Chern-normal insulator phase transitions. At finite temperatures, thermal averaging or dephasing to mixed states results in a deviation from an integer Chern number. We have evaluated the thermal Uhlmann number and showed that the CI phase is stable for temperatures $\lesssim 30$ K.
\begin{figure}[t]
\vspace*{-0.0cm}
\begin{center}
\hspace{-0.2cm}\includegraphics[height=3.4cm, width=8.8cm ]{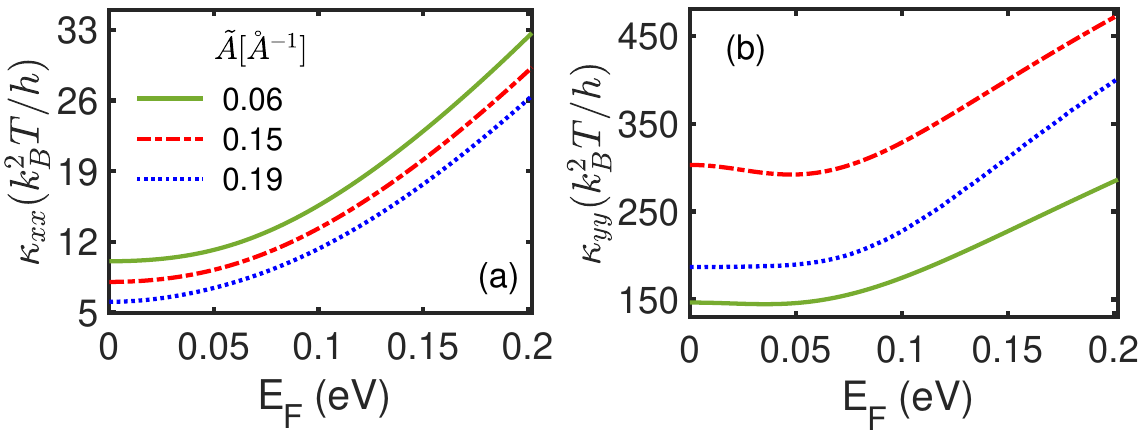}
\end{center}
\vspace*{-0.4cm} \caption{(Colour online) Thermal conductivities (a) $\kappa_{xx}$ and (b) $\kappa_{yy}$ versus $E_F$ for increasing values of the light intensity $\tilde{A}$. The temperature is $T = 300$ K.   }
\label{fig:fig1}%
\end{figure}

The TRS breaking light induces and controls the Berry-phase mediated anomalous thermoelectric responses in SD materials. In particular, we investigated the anomalous Nernst ($\alpha_{xy}$) and thermal Hall ($\kappa_{xy}$) conductivities, and the orbital magnetization $M$. In the BI phase we find that $\alpha_{xy}$ versus $E_F$ exhibits a dip (peak) at the valence (conduction) band edge. The dip-peak profile is reversed when the light intensity drives the system to the CI phase. We also showed that switching the light's circular  polarization, from left to right, induces a sign reversal in $M$, $\sigma_{xy}$, $\alpha_{xy}$ and $\kappa_{xy}$. Physically this allows us to reverse the direction of flow of the transverse thermoelectric charge current and that of the heat current. 

 
We have also investigated the diagonal components of the charge, thermoelectric, and thermal conductivity tensors, $\sigma_{aa}$, $\alpha_{aa}$, and $\kappa_{aa}$, respectively, with $a = x, y$. We find that $\sigma_{yy}$, $\alpha_{yy}$, and $\kappa_{yy}$ depend strongly on the light intensity compared to those in the $x$ direction. Additionally, we showed that the values of $\alpha_{yy}$ and $\kappa_{yy}$ are one and two orders of magnitude larger than $\alpha_{xx}$ and $\kappa_{xx}$, respectively. We also find that the light has opposite effect on $\alpha_{xx}$ and $\alpha_{yy}$; as $\tilde{A}$ increases $|\alpha_{xx}|$ decreases but $|\alpha_{yy}|$ increases. These results may be promising for light-induced caloritronic applications.

\section*{ACKNOWLEDGMENTS}

V.V. and N.N. acknowledge funding from the UK Research and Innovation fund (project reference EP/X02346X/1).

\begin{appendix}

\section{Gap originating from the mass $m$}

A local on-site staggered sublattice potential in honeycomb lattices breaks the equivalence between A and B sites, and therefore also breaks the inversion symmetry \cite{carbot08}. It is expressed as
\begin{equation}
h_s = \sum_{\mathbf{r}_A} E_A c_A^{\dagger} ( \mathbf{r}_A ) c_A ( \mathbf{r}_A ) + \sum_{\mathbf{r}_B} E_B c_B^{\dagger} ( \mathbf{r}_B ) c_B ( \mathbf{r}_B )  ,
\label{eq4300}%
\end{equation}
where $E_A$ and $E_B$ are the orbital energies on sites A and B, respectively. This local perturbation term is nondispersive and its Bloch Hamiltonian in $\mathbf{k}$ space is written as
\begin{equation}
h_s ( \mathbf{k} ) = h_s = m_0 \sigma_0 + m \sigma_z  ,
\label{eq4400}%
\end{equation}
where the first term with $m_0 = (E_A + E_B) / 2$ can be absorbed in the chemical potential $E_F$ and is ignored. The second term with $m = ( E_A - E_B ) / 2$ is parity breaking and opens a gap of size $2 | m |$ at the Dirac points. Such a gap has been demonstrated in epitaxial graphene \cite{giovan}.

A honeycomb lattice can also exhibit SD dispersion when the particle hopping is different along different directions of the lattice, thus breaking the rotational symmetry. An example is graphene deformed by applying a uniaxial strain in one direction \cite{fuch09}. We remark that, for such semi-Dirac honeycomb lattice, direct evidence of the anisotropic transport of polaritons has been reported \cite{real20}. In order to generate a mass for the SD fermions, one possibility is to use deformed graphene on a substrate that makes the A and B sites inequivalent as a result of the interlayer interaction. A possible choice for a substrate is the hexagonal boron nitride (\textit{h}-BN). It has been demonstrated that, for isotropic graphene placed on top of BN, the mass generated is $m \simeq 26$ meV \cite{giovan}. For a SD deformed graphene the interaction with the substrate will be weaker and the expected value of the mass should be lower. However, it has proved challenging to realize the SD dispersion in these systems.

Other honeycomb materials, such as silicene and germanene, were shown to host the SD semimetal phase with the application of structural strain and the addition of oxygen atoms to modify the hopping energies \cite{chem17}. Due to the buckled structure the two sublattice planes of silicene \cite{ez12} are separated by a distance $d = 0.46$ \AA. The application of an electric field $E_z$ perpendicular to the plane generates a staggered sublattice potential with $m = E_z d / 2$. In this case, for $E_z = 10$ meV/{\AA} we estimate that $m \simeq 2.3$ meV and the gap $\simeq 4.6$ meV.

\end{appendix}


\begin{thebibliography}{9}
\bibitem{remark} We remark here that when $m = A_0 =0$, the system is a BI for $\Delta > 0$. As $\Delta$ is varied from positive to negative, the system undergoes a topological phase transition at $\Delta = 0$ to semi-metallic phase ($\Delta < 0$) \cite{fuch09}.
\bibitem{dirac} This is the only real solution for $\mathbf{k}_D$ when $\Delta_1 < 0$. As $\Delta_1 \rightarrow 0$ it yields the positions of the SD nodes with high accuracy.
\bibitem{comment} The quantity $\tilde{A} = e A_0 / \hbar$ characterizes the intensity of the off-resonant light. The light intensity reads $I = e^2 A_0^2 \omega^2 / 8 \pi \hbar \alpha_f$ with $\alpha_f \simeq 1 / 137$ the fine-structure constant. Then, $\tilde{A} = 8 \pi \alpha_f I / \hbar \omega^2$ is an experimentally adjustable parameter for the observation of topological properties with a probe laser \cite{bao22}. Placing the frequency of the off-resonant light in the soft x-ray regime with $\omega = 3 \times 10^{15}$ Hz and choosing a light intensity $I \simeq 19 \times 10^{10}$ W/cm$^2$ gives $\tilde{A} \simeq 0.06$ \AA$^{-1}$ and an electric field strength $E_0 = \hbar \omega \tilde{A} / e = 11 \times 10^8$ V/m. For these values $e A_0 v_y / \hbar \omega \simeq 0.06$, which guarantees the validity of the Floquet approach. The required values of $\tilde{A}$ (and of $E_0$) for the observation of the above phenomena can be significantly lowered if the gap parameter $\Delta$ becomes smaller, for instance due to smaller strain. For $\Delta = -15$ meV we estimate that $E_0 \simeq 5 \times 10^7$ V/m.
\bibitem{mass} The mass $m$ used here is similar to the staggered sublattice potential in honeycomb lattices and could be tuned by adatoms, a substrate, or an out-of plane external electric field.
\bibitem{range} For $\Delta_1 = 0$ the range of $\theta$ is $[- \pi / 2, \pi / 2]$. For $\Delta_1 < 0$ and $| \Delta_1 | \geq q$ it is $[- \pi , \pi]$. For $\Delta_1 > 0$ and $\Delta_1 < q$ it is $[ - \arccos ( \Delta_1 / q ),  \arccos ( \Delta_1 / q )]$.
\bibitem{effective} We remark that the smallness of the ratio $v_x^2 / v_y^2$ is partly a consequence of the relatively large effective mass of few layer black phosphorus. In other SD systems, for example in deformed graphene \cite{fuch09}, the effective mass is approximately half. However, for VO$_2$ systems \cite{pardo09b} it can be larger.
\end{thebibliography}
\end{document}